\documentclass{ws-ijmpd}
\usepackage[super,compress]{cite}

\usepackage{hyperref}

\newcommand{\bra}[1]{\left\langle #1 \right|}
\newcommand{\ket}[1]{\left| #1 \right\rangle}
\newcommand{\braket}[2]{\left\langle {#1{\left| \vphantom{#1 #2} \right.} #2} \right\rangle}

\renewcommand{\epsilon}{\varepsilon}
\renewcommand{\phi}{\varphi}

\begin{document}

\markboth{Qasem Exirifard and Ebrahim Karimi}
{Schr\"odinger equation in a general curved space-time geometry}

%
\catchline{}{}{}{}{}
%

\title{SCHR\"ODINGER EQUATION IN A GENERAL CURVED SPACE-TIME GEOMETRY  }

\author{QASEM EXIRIFARD}

\address{Physics Department, University of Ottawa,\\
	Ottawa, Ontario K1N 6N5, Canada\\
	qexirifa@uottawa.ca}

\author{EBRAHIM KARIMI}

\address{Physics Department, University of Ottawa,\\
	Ottawa, Ontario K1N 6N5, Canada\\
	ekarimi@uottawa.ca}
\maketitle


\begin{abstract}
	We consider relativistic quantum field theory in the presence of an external electric potential in a general curved space-time geometry. We utilise Fermi coordinates adapted to the time-like geodesic to describe the low-energy physics in the laboratory and  calculate the leading correction due to the curvature of the space-time geometry to the Schr\"odinger equation. We then compute the non-vanishing probability of excitation for a hydrogen atom that falls in or is scattered by a general Schwarzschild black hole. The photon that is emitted from the excited state by spontaneous emission extracts energy from the black hole, increases the decay rate of the black hole and adds to the information paradox.    
\end{abstract}


\ccode{PACS numbers: 04.62.+v, 42.50.Ct, 97.60.Lf, 04.70.Dy}

\section{Introduction}
All forces, except gravity, are described in the standard model of elementary particles in a renormalizable quantum field theory framework. It is believed that an extension of the standard model of particle physics to a generally curved space-time geometry describes the interaction between matter and gravity in the quantum realm. The criterion of minimal interaction with the metric and techniques of quantum field theory in curved space-time geometry have been utilized to calculate effects of curved space-time geometry for Hartle \& Hawking observer \cite{Hartle:1976tp} and Unruh observer (vacuum) \cite{Unruh:1976db}. The spontaneous excitation of static atoms due to the non-trivial vacuum has been studied \cite{Yu:2007wv,Zhou:2012eb,Cheng:2019tnk,Zhu:2008bd}. The change in the spectrum of a static hydrogen atom in a curved space-time geometry has been studied \cite{Parker:1982nk,Parker:1980kw}. Here, we develop a framework for computing the excitation of a (non-static)  hydrogen atom that falls freely in a curved space-time geometry. 

We consider a hydrogen atom at its ground state in the asymptotic flat infinity that moves along a time-like geodesic toward a mass distribution. We adapt  Fermi coordinates to describe the curved space-time geometry around the geodesic. In  Fermi coordinates, the metric is the Minkowski metric corrected by the components of Riemann tensor evaluated on the geodesic. Utilizing the effective field approach,  we show that the corrections given by the components of Riemann tensor cause a transition from the ground state. This approach allows us to  provide the leading correction to the Schr\"odinger equation due to the curvature of the space-time geometry. We prove that the leading correction coincides to the Newtonian tidal gravitational force. We also calculate the excitation of a hydrogen atom scattered by a Schwarzschild space-time geometry. 

The work is structured as follows. We first review the systematic extension of the standard model of elementary particles to a generally curved space-time geometry by the minimal interaction criterion is  section \ref{Section2}. We consider the effective action of an electron in the presence of an external electric potential in a general curved space-time geometry in section \ref{Section3}. We review  the construction of the Fermi coordinates adapted to a general time-like geodesic in section \ref{Section4}. The effective action of a trapped electron in Fermi coordinates adapted to the time-like geodesic of the lab is derived in section \ref{Section5}. The low-energy physics of the trapped electron is studied in section \ref{Section5}. The leading corrections by the Riemann tensor of the back geometry to the Schr\"odinger equations are calculated and presented in \eqref{Eq5.15}, where the products of states are defined \eqref{Eq5.19}. After redefining the wave-function by \eqref{Eq5.20}, it is shown that the leading correction by Riemann tensor coincides with the residual Newtonian gravitational potential present in the lab, while inner products between wave-functions are that of the flat space-time geometry. The Schwarzschild space-time geometry is considered in section \ref{Section6}, and the components of the Riemann tensor are calculated in Fermi coordinates for a general time-like geodesic. The hydrogen atom in the flat space-time geometry is reviewed in section \ref{Section7}. The effective gravitational potential felt by the electron when the hydrogen atom moves on a time-like geodesic of a curved space-time geometry is calculated in \eqref{Eq7.15}.

Section \ref{Section8} considers a hydrogen atom radially falling into the Schwarzschild black hole. The atom is assumed to be in its ground state at asymptotic infinity. The transition probability of the hydrogen atom to excited states on the event horizon is calculated in section \ref{Section8}. The rules for transition from $\ket{n,\ell,m}$ to $\ket{n',\ell',m'}$ are provided in \eqref{Eq8.11}. The amplitude for the transition from ground state to the state of $\ket{3,2,0}$, which is the first allowed excited state, is calculated. Figure \ref{fig:fig1} depicts the probability of transition as a function of the Lorentz factor of the hydrogen atom at the asymptotic infinity for a set of Schwarzschild radii. The transition amplitude for ultra-relativistic and classical hydrogen atom is calculated respectively in sub-sections \ref{Section8.1} and \ref{Section8.2}. It is reported that increasing the Lorentz factor of the hydrogen atom enhances the probability of transition. The relativistic enhancement factor is evaluated and presented in section \ref{Section8.3}.

Section \ref{Section9} studies the deflection of a hydrogen atom with arbitrary Lorentz factor at the asymptotic infinity and a general impact parameter from a Schwarzschild black hole. It considers the hydrogen atom in its ground state before the impact with the black hole, and computes the transition amplitude to excited states after the impact. The rules for allowed transitions are derived and presented in \eqref{Eq9.33}. The amplitude for the transition to the first excited states of $\ket{3,2,0}$, $\ket{3,2,\pm1}$ and $\ket{3,2,\pm{2}}$ are computed. It is shown that the behaviour in the ultra-relativistic regime is simplified. It is reported that the amplitude for transition to $\ket{3,2,\pm1}$ increases by increasing the Lorentz factor of the hydrogen atom, while the amplitude of transition to $\ket{3,2,0}$ decreases and that of $\ket{3,2,\pm2}$ approaches a constant value.  

The results are reviewed and remarks are provided in section \ref{Section10}.

\section{Standard model of elementary particles in a curved space-time geometry}
\label{Section2}
The observed matter content of the universe is presented in the standard model of particle physics. The standard model includes leptons, quarks, $W^{\pm}, Z$, $\gamma$ (photon), gluons, and Higgs. It is governed by a renormalizable quantum field theory with spontaneously broken SU(3)$\times$SU(2)$\times$U(1) gauge symmetry. Let $\Psi$ represent all the field content of the standard model. The tree-level action of the standard model can be presented by,
\begin{equation}
	\label{Eq2.1}
	S_{SM}= \int d^4 x\, {\cal L}(\Psi, \partial_\mu \Psi,\eta_{\mu\nu} ),
\end{equation} 
where ${\cal L}$ is the Lagrangian density of the standard model which is a non-linear function of $\Psi$ and its first derivative, and $\eta_{\mu\nu}$ is the Minkowski metric. 

In order to include gravity, we should extend the standard model to quantum field theory in curved space-time geometry \cite{QFT1, QFT2} endowed by a metric. Considering a manifold with a given topology and differential structure,  the assumption of minimal interaction with the metric replaces $\eta_{\mu\nu}$ with an arbitrary metric $g_{\mu\nu}$, resulting in
\begin{equation}
	\label{Eq2.2}
	S_{SM}= \int d^4 x\,\sqrt{-\det g}\, {\cal L}(\Psi, \partial_\mu \Psi,g_{\mu\nu}).
\end{equation} 
The Einstein-Hilbert action that governs the dynamics of the metric is given by,
\begin{equation}
	\label{Eq2.3}
	S_{g}=\frac{c^4}{16 \pi G} \int d^4x\,\sqrt{-\det g}\, R,
\end{equation}
where $G$ is Newton's gravitational constant, and $c$ is the light speed. Therefore, the total action of the theory then is given by,
\begin{equation}
	\label{Eq2.4}
	S = S_g + S_{SM}. 
\end{equation}
The quantum path integral of the theory can be represented by,
\begin{equation}
	\label{Eq2.5}
	Z[J, J^{ab}]= \sum_{TP}
	\frac{\int {\cal D}\!\Psi{\cal D}\!g_{ab} 		e^{-\frac{i}{\hbar}(S+\int d^4x \sqrt{-\det g}\, (J\cdot\Psi+ J^{ab}g_{ab}))}}{{\cal V}_{\cal G }\int {\cal D}\!\Psi {\cal D}\!g_{ab}\,  e^{-\frac{i}{\hbar}(S[\Psi])}},
\end{equation}
where $ \sum_{TP}$ stands for summation over all topologies,  $\int {\cal D}\!\Psi$ and $\int {\cal D}\!g_{ab}$ stand for integration over all field configurations and over all metrics, $J$ represents the source field for $\Psi$, $J^{ab}$ represents the source field for the metric, and ${\cal V}_{\cal G }$ is the volume of the gauge groups. This theory, however, is not perturbatively renormalizable around the flat space-time geometry. In order to be able to utilize the standard methods of the quantum field theory, we ignore the dynamics of the metric and consider quantum field theory of the standard model in a general curved space-time geometry where the quantum path integral of the theory is given by,
\begin{equation}
	\label{Eq2.6}
	Z[J]=\frac{\int {\cal D}\!\Psi e^{-\frac{i}{\hbar}(S_{SM}+\int d^4x \sqrt{-\det g}\, J\cdot\Psi)}}{{\cal V}_{\cal G }\int {\cal D}\!\Psi  e^{-\frac{i}{\hbar}S_{SM}}}.
\end{equation}
This approach is called quantum field theory in a curved space-time geometry. The effective action, $\Gamma$, is defined as the Legendre transformation of logarithm of $Z[J]$: 
\begin{eqnarray}
	Z[J] &\equiv& e^{\frac{i}{\hbar} W[J]},\\
	\Gamma[\Psi_c]&\equiv & W[J]- \int d^4x \sqrt{-\det g} J.\Psi_c,
\end{eqnarray}
where 
\begin{equation}
	\Psi_c \equiv \frac{\delta W[J]}{\delta J}.
\end{equation}
The standard perturbative renormalization approach to the path integral returns the perturbative expansion of the effective action in term of $\hbar$, i.e.,
\begin{equation}
	\label{Eq2.7}
	\Gamma:= \Gamma[\Psi_c]= \Gamma^{(0)}[\Psi_c] +\hbar\,\Gamma^{(1)}[\Psi_c] + \hbar^2\,\Gamma^{(2)}[\Psi_c]  + \ldots.
\end{equation}
The first term coincides with the tree-level action:
\begin{equation}
	\label{Eq2.8}
	\Gamma^{(0)}[\Psi_c]=\int d^4 x\sqrt{-\det g}\, {\cal L}(\Psi_c, \partial_\mu \Psi_c,g_{\mu\nu} ),
\end{equation} 
and the sub-leading corrections can be computed. In high densities,  the sub-leading corrections can be important \cite{Latorre:2017uve}. We consider low-energy densities where the sub-leading corrections can be ignored. 

The effective action includes both the classical and quantum effects. The classical effects are those that can be reproduced by the motion of point-like particles  along geodesics or by classical fields. The rest are quantum. Hawking effect \cite{Hawking:1974rv} is a known quantum effect due to curved space-time geometry that was initially discovered by studying quantum field theory in black hole space-time geometry within the approach of the operator product expansion method. However, the effective field theory approach provides a simpler method to compute it \cite{Damour:1976jd,Vieira:2014waa}. We would like to utilize the effective field theory approach to find how a curved space-time geometry affects quantum mechanics and alters the Schr\"odinger equation.

\section{Massive charged scalar field in the presence of an electric potential in curved space-time geometry}
\label{Section3}
The approaches to obtain an effective low-energy description of a quantum particle in curved space-time include the expansion of the Dirac equation in the generalised Fermi coordinates along a reference world-line \cite{Ito:2020xvp,Perche:2020lzz}, or less-explicit geometric expansion of the Klein--Gordon equation  to obtain an effective Schr\"odinger equation with ‘relativistic corrections’ \cite{Schwartz:2018pnh}. In this section, we treat the Schr\"odinger equation as the low-energy limit of the Klein--Gordon equation and calculate the geometric corrections to the Schr\"odinger equation.  In so doing, we first would like to justify the consideration of the Klein--Gordon equation in describing the behavior of an electron in a curved spaced-time geometry where physics is governed by the Dirac equation: 
\begin{equation}
	\label{Eq3.1}
	(i \gamma^\mu {\cal D}_\mu-m) \psi =0,
\end{equation}
where, 
\begin{equation}
	\label{Eq3.2}
	{\cal D}_\mu = \partial_\mu - \Gamma_\mu + i e A_\mu.
\end{equation}
$\gamma_\mu$ are the generalized gamma matrices satisfying the covariant Clifford algebra, 
\begin{equation}
	\label{Eq3.3}
	\gamma_{\mu}\gamma_\nu + \gamma_\nu \gamma_\mu = -2 g _{\mu\nu},
\end{equation}
while $\Gamma_\mu$ is the spinorial affine connection,  $A_\mu$ is the electromagnetic four-vector potential, and $e$ is the electric charge of the fermion. The modified Klein--Gordon equation obtained by squaring the operator in the Dirac equation,  first found by Schr\"odinger  as cited by Pollock \cite{Pollock:2010zz}, is given by,
\begin{equation}
	\label{Eq3.4}
	\left(\frac{1}{\sqrt{-\det g}}\, {\cal D}_\mu \left(\sqrt{-\det g}\, g^{\mu\nu}{\cal D}_\nu\right)- \frac{1}{4} R + \frac{i e}{2} F_{\mu\nu} s^{\mu\nu} - m^2\right)\Psi=0.
\end{equation}
Here, $R$ is the Ricci scalar, and $F_{\mu\nu}$ is the field strength of $A_\mu$, i.e.,
\begin{equation}
	\label{Eq3.5}
	F_{\mu\nu}=\partial_\mu A_\nu -\partial_\nu A_\mu.
\end{equation}
For Ricci flat space-time geometries, when the spin of the fermion (electron) can be neglected,  the spinor $\Psi$ can be replaced by the scalar field $\phi$, ${\cal D}_\mu$ converts to,
\begin{equation}
	\label{Eq3.6}
	{\cal D}_\mu = \partial_\mu +i e A_\mu, 
\end{equation}
and \eqref{Eq3.4} can be approximated to,
\begin{equation}
	\label{Eq3.7}
	\left(\frac{1}{\sqrt{-\det g}}\, {\cal D}_\mu \left(\sqrt{-\det g}\, g^{\mu\nu}{\cal D}_\nu\right) - m^2\right)\phi=0. 	
\end{equation}
Let it be emphasised that, in obtaining \eqref{Eq3.7}, we have assumed that the electron is not relativistic with respect to the hydrogen atom and that the frame-dragging effect is smaller than the Newtonian potential. So, the gravitomagnetic interaction between the electron's spin or angular momentum can be consistently neglected at the leading order approximation.  
Eq. \eqref{Eq3.7} is the variation of,
\begin{eqnarray}
	\label{Eq3.8}
	\Gamma[\phi]&=& \frac{1}{2}\int d^4 x \sqrt{-\det g}\, \left(g^{\mu\nu} {\cal D}_\mu \phi ({\cal D}_\nu\phi)^* + m^2 \phi \phi^*\right),
\end{eqnarray}
with respect to $\phi^*$.  Note that $\phi^*$ is the complex conjugate of  $\phi$, $\hbar$ and $c$ are set to 1 ($\hbar=c=1$), and the metric has three positive eigenvalues. Equation \eqref{Eq3.8} is known as the scalar approximation to the electron action. Now, let us consider an electric potential, where the four-potential is given by,
\begin{eqnarray}
	\label{Eq3.9}
	A_\mu = (V,0,0,0).
\end{eqnarray}
For a weak electric field, one can ignore the quadratic term in $V$ in \eqref{Eq3.8} and thus obtain,
\begin{eqnarray}
	\label{Eq3.10}
	\Gamma[\phi]= \frac{1}{2}\int d^4 x \sqrt{-\det g} &&\left(g^{\mu\nu} {\partial}_\mu \phi {\partial}_\nu\phi^* - i e g^{00}V\phi^*\partial_0 \phi \right.\\ \nonumber &&\left. + i e g^{00}V\phi\partial_0 \phi^*+  m^2 \phi \phi^*\right),
\end{eqnarray}
The low-energy physics is described by, 
	\begin{eqnarray} 	\label{Eq3.11}
		\phi &=& e^{-im t} \Psi(t,\vec{x}),\\
		|\partial_t \Psi| &\ll& |m \Psi |.
	\end{eqnarray}
Therefore, we can substitute $\partial_0 \phi$ with $-im \phi$ and simplify \eqref{Eq3.10} to,
\begin{eqnarray}
	\label{Eq3.12}
	\Gamma[\phi]&=& \frac{1}{2}\int d^4 x \sqrt{-\det g} \left(g^{\mu\nu} {\partial}_\mu \phi {\partial}_\nu\phi^* - 2 e m g^{00} V\phi \phi^* +  m^2 \phi \phi^*\right),
\end{eqnarray}
which we refer to as the action of a massive charged particle in a curved space-time geometry under the external electric potential $V.$ 

In the case of flat space-time geometry, $g^{\mu\nu}={{diag}}(-1,1,1,1)$, the variation of \eqref{Eq3.12} with respect to $\phi^*$ yields, 
\begin{equation}
	\label{Eq3.13}
	\left(\nabla^a \nabla_a-m^2\right) \phi = \left(-\partial^2_t + \partial^a \partial_a-m^2 - 2 e m V\right) \phi.
\end{equation} 
Utilizing \eqref{Eq3.11} in \eqref{Eq3.13} enables us to ignore the second derivative of $\Psi$ with respect to time and obtain,
\begin{equation}
	\label{Eq3.14}
	\left(2 i m \partial_t  + \partial_a \partial^a- 2 e m V\right) \Psi = 0 ,
\end{equation} 
which can be rewritten into the following form,
\begin{eqnarray}
	\label{Eq3.15}
	\left(-\frac{\hbar^2}{2m} \nabla^2 +e V\right) \Psi(t,\vec{x}) =  i  \hbar\, \partial_t \Psi(t,\vec{x}).
\end{eqnarray}
Here, we recovered the $\hbar$, and $\nabla^2 = \partial^a \partial_a$. Equation \eqref{Eq3.15} is the Schr\"odinger equation for a particle of mass $m$ and charge $e$ in the presence of electric potential $V.$ The low-energy physics of \eqref{Eq3.12} for a general metric includes the corrections to the Schr\"odinger equation due to the curvature of the space-time geometry. In order to extract the corrections, we should first fix the general covariance symmetry of the theory by appropriately choosing the coordinates. 

\subsection{Fermi coordinates adapted to time-like geodesics}
\label{Section4}
We can choose the rest frame of the object producing the potential; we call this frame the lab frame. The lab is considered to move along a time-like geodesic. We choose Fermi coordinates to describe the space-time geometry in the lab. The expansion of Fermi coordinates  adapted to the time-like geodesic $\gamma$, up to the quadratic transverse directions, are given by Manasse and Misner \cite{Manasse:1963zz}:\footnote{Fermi coordinates adapted to null geodesics is provided by Matthias Blau, Denis Frank, Sebastian Weiss \cite{Blau:2006ar}}
\begin{eqnarray}
	\label{Eq4.1}
	ds^2 &=& c^2 (dx^0)^2 \left(- 1+ R_{0l0m} x^l x^m\right) +\nonumber\\&+& \frac{4}{3} R_{0 l i m} x^l x^m dx^0 dx^i+ dx^i dx^j \left(\delta_{ij}+ \frac{1}{3} R_{iljm} x^l x^m\right) \nonumber\\
	&+&O(x^l x^m x^n)\,,
\end{eqnarray}
where $R_{\mu\alpha\beta\nu}$ represents the components of the Riemann tensor computed along the time-like geodesics, and $x^i$ are the spatial transverse directions to the time-like geodesic. Note that $x^0$ is the proper time in the lab, i.e., $x^0= \tau$. We choose units such that $c=1$.  Therefore, we can write the following systematic expansion series for the metric, 
\begin{eqnarray}
	\label{Eq4.2}
	g_{\mu\nu}=g_{\mu\nu}^{(0)}+ \epsilon g_{\mu\nu}^{(1)}+O(\epsilon^2)\,,
\end{eqnarray}
where $\epsilon$ is the systematic parameter of the series, and, 
\begin{eqnarray}
	\label{Eq4.3}
	g_{\mu\nu}^{(0)} = \eta_{\mu\nu}\,,
\end{eqnarray}
where $\eta_{\mu\nu}$ stands for the Minkowski metric,
\begin{eqnarray}
	\label{Eq4.4}
	\eta_{\mu\nu} dx^\mu dx^\nu = - dt^2 + dx^i dx^i\,,
\end{eqnarray}
and, 
\begin{subequations}
    \label{Eq32Revision}
	\begin{eqnarray}
		\label{Eq4.5}
		g_{00}^{(1)} &=& R_{0l0m} x^l x^m \,,\\
		g_{0i}^{(1)} &=& \frac{2}{3} R_{0 l i m} x^l x^m\,,\\
		g_{ij}^{(1)} &=& \frac{1}{3} R_{iljm} x^l x^m\,.
	\end{eqnarray}
\end{subequations}

``The expansion of the metric in Fermi coordinates in eq. \eqref{Eq4.1} is directly cited from Manasse and Misner's work\cite{Manasse:1963zz} with a non-standard sign convention for the Riemann tensor: ${}^\text{(Manasse-Misner)} R_{\nu~\rho\sigma}^{~\mu}= {}^\text{(MTW)} R_{~\nu\rho\sigma}^{\mu}$, where ${}^\text{(MTW)} R_{~\nu\rho\sigma}^{\mu}$ is the standard sign convention in GR textbook by Misner, Thorne and Wheeler \cite{Misner:1973prb} which is the most widely used sign convention nowadays"\footnote{We thank referee 2 for pointing this to us.}. We observe that if we set $\epsilon=-1$, where $\epsilon$ is the systematic parameter of the expansion, then the Manasse and Misner convention is mapped to nowadays' standard convention. We shall use the nowadays' convention and set $\epsilon=-1$ at the end of the computation.
In the following, we would like to compute the determinant and the inverse of the metric. The determinant of the metric, i.e., $\det g$, is,
\begin{equation}
	\label{Eq4.6}
	-\det g = 1 + \epsilon \left(- R_{0 l 0 m} + \frac{1}{3} R_{k l n m} \delta^{kn} \right)x^l x^ m + O(\epsilon^2)\,.
\end{equation}
We are interested in the vacuum solutions of the Einstein equations where the Ricci tensor vanishes. The components of the Ricci tensor evaluated on the geodesic are, 
\begin{equation}
	\label{Eq4.7}
	R_{lm}= \eta^{\mu\nu} R_{l \mu m \nu}=\delta^{kn}  R_{k l n m}-R_{0l0m} = 0 \,.
\end{equation}
This can be used to simplify Eq.~\eqref{Eq4.6}:
\begin{equation}
	\label{Eq4.8}
	-\det g = 1 - \frac{2\epsilon }{3}  R_{0 l 0 m} x^l x^ m  + O(\epsilon^2)\,.
\end{equation}
Therefore, if we represent $g = -\det g$, then, 
\begin{eqnarray}
	\label{Eq4.9}
	g &=& 1 + \epsilon g^{(1)},\\
	\label{Eq4.10}
	g^{(1)}&=& - \frac{2 }{3}  R_{0 l 0 m} x^l x^ m\, .
\end{eqnarray}
In order to compute the expansion series for the inverse of the metric,  the definition of the inverse metric ($g^{\mu\nu} g_{\nu\xi} = \delta^\mu_{~\xi}  $) is utilized to write down,
\begin{equation}
	\label{Eq4.11}
	(\eta^{\mu\nu} + \epsilon g^{(1)\mu\nu}) (\eta_{\nu\xi}+\epsilon g^{(1)}_{\nu\xi}) = \delta^\mu_{~\xi} + O(\epsilon^2)\,,
\end{equation}
which implies, 
\begin{equation}
	\label{Eq4.12}
	g^{(1)\mu\nu} = - \eta^{\alpha\mu} \eta^{\beta\nu} g^{(1)}_{\alpha\beta}\,.
\end{equation}
Since $\eta^{\mu\nu}$ is diagonal, and its diagonal values are $\pm 1$, one gets:
\begin{equation}
	\label{Eq4.13}
	g^{(1)\mu\nu} = - g^{(1)}_{\mu\nu}\,.
\end{equation}
Therefore, 
%
\begin{subequations}
\label{Eq4.14}
	\begin{eqnarray}
		g^{(1)00} &=& -R_{0l0m} x^l x^m \,,\\
		g^{(1)0i} &=& -\frac{2}{3} R_{0 l i m} x^l x^m\,,\\
		g^{(1)ij} &=& -\frac{1}{3} R_{iljm} x^l x^m\,.
	\end{eqnarray}
\end{subequations}
%

\subsection{Correction by the Riemann tensor to the Schr\"odinger equation}
\label{Section5}
The quantum field theory of a massive scalar particle in the background potential $V$ in a general curved space-time geometry was presented in \eqref{Eq3.12}.  The functional variation of  \eqref{Eq3.12} with respect to $\phi$ gives its equation of motion:
\begin{equation}
	\label{Eq5.1}
	\left(\frac{1}{\sqrt{-\det g}}\partial_\mu \left(\sqrt{-\det g}g^{\mu\nu} \partial_\nu\right)-   m^2 + 2 e m  V g^{00}\right) \phi = 0	.
\end{equation}
We assume that the potential is produced by a massive entity; the entity can be a proton or the lab. We choose Fermi coordinates along the time-like geodesic of the entity. Section \ref{Section4} has presented the components of the metric along a time-like geodesic in Fermi coordinates up to quadratic order in the transverse directions to the geodesic. This allows for the perturbative $\epsilon$ expansion series for the metric, its determinant and the inverse:
%
	\begin{eqnarray}
		\label{Eq5.2}
		g_{\mu\nu} &=& \eta_{\mu\nu}+ \epsilon g^{(1)}_{\mu\nu} + O(\epsilon^2),\\\nonumber
		g^{\mu\nu} &=& \eta^{\mu\nu}+ \epsilon g^{(1)\mu\nu} + O(\epsilon^2),\\ \nonumber
		g &=& 1+ \epsilon g^{(1)} + O(\epsilon^2),
	\end{eqnarray}
%
where $g$ represents ``$-\det g$''. In Eq. \eqref{Eq32Revision}, $g^{(1)}_{\mu\nu}$ are expressed in terms of the components of the Riemann tensor evaluated on the geodesic. Equation \eqref{Eq4.14} shows the corresponding expression for $g^{(1)\mu\nu}$, and the expression for $g^{(1)}$ is shown in \eqref{Eq4.10}. 
Utilizing \eqref{Eq5.2} gives, 
\begin{eqnarray}
	\label{Eq5.3}
	\left(\nabla^a \nabla_a-m^2- 2 e m V\right) \phi + \frac{\epsilon}{2} \eta^{\mu\nu} \partial_\mu g^{(1)} \partial_\nu \phi &+\\ \epsilon \partial_\mu \left(g^{(1)\mu\nu} \partial_\nu \phi\right)  + 2\epsilon e m V g^{(1)00}\phi=  O(\epsilon^2)  \,,
\end{eqnarray}
where $ \nabla_a \nabla^a$ represents the d'Alembert operator in flat space-time geometry,
\begin{equation}
	\label{Eq5.4}
	\nabla^a \nabla_a=  \eta^{\mu\nu} \partial_\mu \partial_\nu= - \partial_t^2 + \nabla^2 \,,
\end{equation}
and $ \nabla^2$ is the Laplace operator in the spatial directions transverse to the geodesic. Since the potential $V$ is small, we can omit the term that includes both $V$ and $\epsilon$ to obtain,
\begin{eqnarray}
	\label{Eq5.5}
	\left(\nabla_a \nabla^a-m^2- 2 e m V\right) \phi +\frac{\epsilon}{2} \eta^{\mu\nu} \partial_\mu g^{(1)} \partial_\nu \phi+ \epsilon \partial_\mu \left(g^{(1)\mu\nu} \partial_\nu \phi\right)  &+& \nonumber \\
	    O(\epsilon^2, \epsilon V)=0\,.
\end{eqnarray}
We also would like to consider the low-energy physics where it holds \eqref{Eq3.11}. Utilizing \eqref{Eq3.11}, and neglecting the second partial derivative of $\Psi$ with respect to time, yields,
\begin{eqnarray}
	\label{Eq5.6}
	\left(\nabla^a\nabla_a-m^2- 2 e m V\right) \phi&=& e^{-imt}\left(\nabla^a\nabla_a+ 2 i m \partial_0 - 2 e m V\right) \Psi \nonumber \\&+&O( \partial_0^2 \Psi),
\end{eqnarray}
which is the first term on the left hand side of \eqref{Eq5.4}. In low-energy physics, all derivatives of $\Psi$ are small compared to the mass:
\begin{equation}
	\label{Eq5.7}
	\forall \mu \quad|\partial_\mu \Psi | \ll m |\Psi|.
\end{equation} 
So the dominant linear correction by the Riemann curvature to the  Schr\"odinger equation includes $\Psi$ but not its derivatives. 

We would like to express  \eqref{Eq5.5} in terms of  $\Psi $. To this aim, notice that the second term in the left hand side of  \eqref{Eq5.5} is given by,
\begin{eqnarray}
	\label{Eq5.8}
	\frac{1}{2} \eta^{\mu\nu} \partial_\mu g^{(1)} \partial_\nu \phi&=&  -\frac{1}{2} \partial_0 g^{(1)} \partial_0 \phi +\frac{1}{2} \partial_a g^{(1)} \partial^a \phi \nonumber\\
	&=& e^{-imt}\frac{im}{2} \partial_0 g^{(1)}  \Psi + O(\partial \Psi) .
\end{eqnarray}
The last term on the left hand side of \eqref{Eq5.5} can be written as,
\begin{eqnarray}
	\label{Eq5.9}
	\partial_\mu \left(g^{(1)\mu\nu} \partial_\nu \phi\right)= g^{(1)\mu\nu}\partial_\mu\partial_\nu \phi+	\partial_\mu g^{(1)\mu\nu} \partial_\nu \phi.
\end{eqnarray}
In order to express \eqref{Eq5.9} in terms of $\Psi$, we first expand its superscripts in terms of $t$ and $x^a$:
\begin{eqnarray}
	\label{Eq5.10}
	\partial_\mu \left(g^{(1)\mu\nu} \partial_\nu \phi\right)&=&g^{(1)00} \partial_0^2 \phi + 2 g^{(1)0a}\partial_0 \partial_a \phi + g^{(1)ab}\partial_a \partial_b \phi \nonumber\\
	&+&\partial_0 g^{(1)00} \partial_0 \phi + \partial_0 g^{(1)0a}\partial_a \phi + \partial_a g^{(1)0a}\partial_0 \phi \nonumber \\&+& \partial_a g^{(1)ab}\partial_b \phi .
\end{eqnarray}
Note, it holds that,
\begin{eqnarray}
	\label{Eq5.11}
	\partial^i g^{(1)}_{0i}= \frac{2 }{3} R_{0iim} x^m,
\end{eqnarray}
where \eqref{Eq4.14} is employed. However, $R_{0iim}$ is proportional to the $0m$ component of the Ricci tensor evaluated on the geodesic. Therefore, Eq.~\eqref{Eq5.11} vanishes. 

Utilizing \eqref{Eq3.11} in \eqref{Eq5.10} yields:
\begin{eqnarray}
	\label{Eq5.12}
	\partial_\mu \left(g^{(1)\mu\nu} \partial_\nu \phi\right)&=&e^{-imt}\left(-m^2 g^{(1)00}  \Psi- i m \partial_0 g^{(1)00}  \Psi\right) + O(\partial \Psi).
\end{eqnarray}
Utilizing \eqref{Eq5.6}, \eqref{Eq5.8}, \eqref{Eq5.12}  in \eqref{Eq5.5} yields, 
\begin{eqnarray}
	\label{Eq5.13}
	\left(\nabla^2+ 2 i m \partial_0 - 2 e m V\right) \Psi +\epsilon \left(\frac{im}{2} \partial_0 g^{(1)} -m^2 g^{(1)00} \right)  \Psi &+& \nonumber \\ O(\epsilon^2, \epsilon\partial\Psi,\partial_0^2 \Psi, \epsilon V) = 0.
\end{eqnarray}
Dividing both sides of \eqref{Eq5.13} by ``$-2m$'' yields:
\begin{eqnarray}
	\label{Eq5.14}
	\left(-\frac{1}{2m}\nabla^2-  i  \partial_0 +  e V\right) \Psi +\epsilon \left(\frac{1}{2}m g^{(1)00}  - \frac{i}{4} \partial_0 g^{(1)} \right)\Psi&+& \nonumber\\   O(\epsilon^2, \epsilon\partial\Psi,\partial_0^2 \Psi, \epsilon V) =0.
\end{eqnarray}
Finally, one may use \eqref{Eq4.14} and \eqref{Eq4.10} to express $g^{(1)00}$ and  $g^{(1)}$ explicitly in terms of the components of the Riemann tensor evaluated on the time-like geodesic of the lab, and thus, we have,
\begin{eqnarray}
	\label{Eq5.15}
	i  \partial_0 \Psi &=& \left(-\frac{1}{2m}\nabla^2 + e V\right) \Psi
	\nonumber \\
	&-&\epsilon\left(\frac{1}{2}m R_{0a0b} x^a x^b 
	-\frac{i}{6} x^a x^b  \partial_0 R_{0a0b} 
	\right)\Psi \nonumber\\
	&+& O(\epsilon^2, \epsilon\partial\Psi,\partial_0^2 \Psi, \epsilon V),
\end{eqnarray}
where $\hbar$ and $c$ are set to $1$. The curvature of the space-time geometry modifies the electrodynamics' equation. The modification causes $V$  to be corrected from it value in flat space-time geometry. In other words, $V$ also has a perturbation in $\epsilon$: $V = V^{(0)} + \epsilon V^{(1)} + O(\epsilon^2)$.
It is noticed that dimensional analysis requires $V^{(1)}$ to be proportional to $V^{(0)}$. But Eq. \eqref{Eq5.15} neglects terms at order $\epsilon V^{(0)}$. Terms at order of $\epsilon V^{(0)}$ are proportional to the binding energy of the electron ($13.6 eV$) while  \eqref{Eq5.15} presents terms that are at the order of the rest mass of the electron ($0.51 MeV$). So it is legitimate to ignore terms at the order of $\epsilon V^{(0)}$  and keep \eqref{Eq5.15} as the leading correction.

  Equation \eqref{Eq5.15} presents the linear correction due to the Riemann curvature of the space-time geometry to the Schr\"odinger equation.

Let it be emphasized that the volume element of a curved space-time geometry is,
\begin{equation}
	\label{Eq5.16}
	{d \cal V}= d^4 x\,\sqrt{-\det g}.
\end{equation}
In Fermi coordinates, it can be written as, 
\begin{equation}
	\label{Eq5.17}
	{d \cal V}= d\tau\, \sqrt{-\det g}\, d^3x,
\end{equation}
where $\tau$ is the proper time in the lab, and $d^3x$ are the spatial coordinates in Fermi coordinates. This enforces us to consider the volume element in $x^1, x^2$, and $x^3$ coordinates by,
\begin{equation}
	\label{Eq5.18}
	{d V}=  \sqrt{-\det g}\, d^3x.
\end{equation}
The amplitude for probability of transition from the state of $\Psi^1$ to $\Psi^2$ is then given by,
\begin{equation}
	\label{Eq5.19}
	{\cal P}(\Psi^1,\Psi^2) =\int dV\,\Psi^{1*} \Psi^2 =  \int d^3x\, \sqrt{-\det g}\, \Psi^{1*} \Psi^2.
\end{equation}
It is convenient to define the physical wave function by, 
\begin{equation}
	\label{Eq5.20}
	\Psi_{Phys}= (-\det g)^{\frac{1}{4}} \Psi.
\end{equation}
The transition amplitude between physical states resembles the transitional amplitude in flat space-time geometry: 
\begin{equation}	
	\label{Eq5.21}
	{\cal P}(\Psi^1_{Phy},\Psi^2_{Phy}) =  \int d^3x\,  \Psi^{1*}_{{Phy}} \Psi^2_{{Phy}}.
\end{equation}
Utilizing \eqref{Eq4.8} and \eqref{Eq5.20} in \eqref{Eq5.15} gives the equation of motion for the physical wave-function:
\begin{eqnarray}
	\label{Eq5.22}
	i  \partial_0 \Psi_{Phy} &=& \left(-\frac{1}{2m}\nabla^2 + e V-\epsilon\frac{m}{2} R_{0a0b} x^a x^b \right) \Psi_{Phy}
	\nonumber \\
	&+& O(\epsilon^2, \epsilon\partial\Psi,\partial_0^2 \Psi, \epsilon V).
\end{eqnarray}
This is the correction to the Schr\"odinger equation in curved space-time geometry where the amplitude of probability is defined by \eqref{Eq5.21}. The procedure of defining the physical wave function removes the imaginary part of the effective potential. We expect the imaginary part of the effective potential in \eqref{Eq5.15} to vanish for physical states to all orders of the approximation because the electric charge is conserved at the tree-level action and the measure of path integral in \eqref{Eq2.6} is invariant with respect to the U(1) symmetry. There exists no anomaly at the quantum level violating the electric charge.

Let us set $\epsilon=-1$, drop the superscript of `Phy', and define,
\begin{equation}	
	\label{Eq5.23}
	V_{eff}= + \frac{m}{2} R_{0 a 0 b} x^a x^b,
\end{equation}
then the Schr\"odinger equation is given by,
\begin{eqnarray}	
	\label{Eq5.24}
	i \hbar  \partial_0 \Psi &=& \left(-\frac{\hbar^2}{2m}\nabla^2 + e V+ V_{eff}\right) \Psi,
\end{eqnarray}  
where $\hbar$ is recovered. Here, the perturbative potential $V_{eff}$ is the residual Newtonian gravitational potential present in the lab, and the amplitude of probability is that of flat space-time geometry.  

Note that $V_{eff}$ is a perturbative time-dependent potential in the lab.  The Hamiltonian of the system can be written as, 
\begin{equation}	
	\label{Eq5.25}
	H = H_0 + V_{eff}(\tau),
\end{equation}
where, 
\begin{equation}	
	\label{Eq5.26}
	H_0 = \frac{P^2}{2m} +e V.
\end{equation}
So the system, initially in the unperturbed eigenstate energy $\ket{\alpha}=\ket{\Psi(\tau_0)}$ by the perturbation, can go into the energy eigenstate  $\ket{\beta}$. Standard perturbation theory gives the transition amplitude to the first order,
\begin{equation}	
	\label{Eq5.27}
	A_{\alpha\beta}= -\frac{i}{\hbar} \int_{\tau_0}^\tau d\tau_1 \bra{\beta}V_{eff}\ket{\alpha} e^{-\frac{i}{\hbar}\left(E_\alpha-E_\beta\right)(\tau_1-\tau_0)},
\end{equation}
where $E_\alpha$ and $E_\beta$ are the energy of the states $\ket{\alpha}$ and $\ket{\beta}$, 
\begin{eqnarray}	
	\label{Eq5.28}
	H_0 \ket{\alpha} &=& E_\alpha \ket{\alpha},\\
	\label{Eq5.29}
	H_0 \ket{\beta} &=& E_\beta \ket{\beta},
\end{eqnarray}
and in the Dirac notation, it holds that,
\begin{equation}
	\label{Eq5.30}
	\bra{\beta}V_{eff}\ket{\alpha} = \int d^3x \int d^3x'\, \braket{\beta}{x} \braket{x}{V_{{eff}}(\tau)|x'}  \braket{x'}{\alpha}.
\end{equation}
The residual Newtonian gravitational potential holds,
\begin{equation}
	\label{Eq5.31}
	\braket{x}{V_{{eff}}(\tau)|x'}= \delta^3(\vec{x}-\vec{x}') V_{{eff}}(\tau,x),
\end{equation}
which simplifies \eqref{Eq5.30} to,
\begin{equation}
	\label{Eq5.32}
	\bra{\beta}V_{{eff}}\ket{\alpha} = \int d^3x  \braket{\beta}{x} V_{{eff}}(\tau)  \braket{x}{\alpha},
\end{equation}
where the dependency of $V_{{eff}}(\tau)$ on $x$ is understood. 
\section{The Schwarzschild space-time geometry}
\label{Section6}
In a space-time endowed with the metric $g_{\mu\nu}$, a geodesic $x^{\mu}(\tau)$ can be obtained from an effective action, i.e., 
%
	\begin{eqnarray}\label{Eq6.1}
		S&=& \int d\tau\, {\cal L}\,,\\
		{\cal L}&=& g_{\mu\nu} \dot{x}^\mu \dot{x}^\nu\,,
	\end{eqnarray}
%
where $\tau$ is an affine parameter. For the Schwarzschild black hole in the standard coordinates, this is,
\begin{equation}
	\label{Eq6.2}
	{\cal L} = -\left(1-\frac{r_s}{r}\right)\dot{t}^2+ \frac{\dot{r}^2}{1-\frac{r_s}{r}}  + r^2 \left(\dot{\theta}^2+ \sin^2 \theta \dot{\phi}^2\right)\,,
\end{equation}
where $r_s ={2 G_N M_{\bullet}}$, $M_{\bullet}$ is the mass of the black hole, and $c=1$. We choose the units such that, 
\begin{equation}
	\label{Eq6.3}
	r_s=\hbar=c=1. 
\end{equation}
Due to the spherical symmetry, without loss of generality, we can choose the equatorial plane, i.e.,
%
	\begin{eqnarray}	\label{Eq6.4}
		\theta&=& \frac{\pi}{2}, \\
		\dot{\theta}&=&0,
	\end{eqnarray}
%
to describe any given geodesic at all times. The cyclic variables of $\phi$ and $t$ lead to invariant quantities:
\begin{eqnarray}
	\label{Eq6.5}
	\frac{\partial {\cal L}}{\partial \phi} = 0 &\to& r^2 \dot{\phi} = l \,,\\
	\label{Eq6.6}
	\frac{\partial {\cal L}}{\partial t} = 0 &\to& \left(1-\frac{1}{r}\right) \dot{t} = E\,.
\end{eqnarray}
Due to the form of the Lagrangian, its Legendre transformation, which is the Lagrangian itself, is invariant.  We set, 
\begin{equation}
	\label{Eq6.7}
	{\cal L} = -\left(1-\frac{1}{r}\right)\dot{t}^2+ \frac{\dot{r}^2}{1-\frac{1}{r}}  + r^2 \ \dot{\phi}^2 = -1\,.
\end{equation}
  The non-zero components of the Riemann tensor in nowadays' standard convention on the geodesic ($\theta=\frac{\pi}{2}$) in the standard spherical coordinates  are:
%
\begin{subequations}	\label{Eq6.8}
	\begin{eqnarray}
		R_{trtr}&=&-\frac{1}{r^3},\\ 
		R_{\theta\phi\theta\phi}&=& r,\\ 
		R_{t\theta t\theta} & =& R_{t\phi t\phi} = \frac{r-1}{2r^2},\\
		R_{r\theta r\theta} &=&R_{r\phi r\phi} =  -\frac{1}{2\left(r-1\right)}.
	\end{eqnarray}
\end{subequations}
%
As a consistency check, we have checked that the Ricci tensor constructed out from \eqref{Eq6.8} vanishes. The coordinate-independent representation of the Riemann tensor, therefore, follows: 
\begin{eqnarray}
	\label{Eq6.9}
	R&=&-\frac{1}{r^3} \left(dt\!\wedge\!dr\right)\otimes \left(dt\!\wedge\!dr\right)
	+r\left(d\theta\!\wedge\!d\phi\right)\otimes\left(d\theta\!\wedge\!d\phi\right)
	\nonumber\\
	&+&\frac{r-1}{2r^2} (dt\!\wedge\!d\phi)\otimes\left(dt\!\wedge\!d\phi\right)+\frac{r-1}{2r^2} \left(dt\!\wedge\!d\theta\right)\otimes\left(dt\!\wedge\!d\theta\right) 
	\nonumber\\
	&-&\frac{1}{2(r-1)} (dr\!\wedge\!d\phi)\otimes(dr\!\wedge\!d\phi)-\frac{1}{2(r-1)} (dr\!\wedge\!d\theta)\otimes(dr\!\wedge\!d\theta)\,.
\end{eqnarray}

\subsection{General time-like geodesic reaching asymptotic infinity}
\label{Section6.1}
In this section, we consider a time-like geodesic that reaches the asymptotic infinity and $l\neq 0$. We assume that the absolute value of the velocity of the lab at the asymptotic infinity is $v$.  Equation \eqref{Eq6.5} implies that, for a finite value of  $l$, the velocity of the lab is radial at the asymptotic infinity because $r \dot{\phi}$ vanishes at the asymptotic infinity. So it  holds that,
\begin{eqnarray}
	\label{Eq6.10}
	\dot{r}|_{r=\infty} &=&- \gamma v,\\
	\label{Eq6.11}
	\dot{t}|_{r=\infty} &=& \gamma ,
\end{eqnarray}
where $\gamma$ is the Lorentz factor and given by,
\begin{eqnarray}
	\label{Eq6.12}
	\gamma= \frac{1}{\sqrt{1-v^2}}.
\end{eqnarray}
Equation \eqref{Eq6.6} implies $E=\gamma$ and yields, 
\begin{equation}
	\label{Eq6.13}
	\dot{t} = \frac{\gamma r}{r-1},
\end{equation}
that can be utilized in \eqref{Eq6.7} to obtain, 
\begin{equation}
	\label{Eq6.14}
	\frac{r (\dot{r}^2- \gamma^2)}{r-1}   = -1 -\frac{l^2}{r^2},
\end{equation}
where \eqref{Eq6.5} is used too, and which can be solved for $\dot{r}$, 
\begin{equation}
	\label{Eq6.15}
	\dot{r}= \pm\sqrt{\gamma^2 -\frac{(r-1)(r^2+l^2)}{r^3} }.
\end{equation}
Here, the minus sign ($-$) is for the lab moving toward the black hole and the plus sign ($+$) is for the lab moving away from the black hole. The tangent to the geodesic is,
\begin{eqnarray}
	\label{Eq6.16}
	\dot{\vec{\gamma}} &=&  h  \dot{t}\, \hat{e}_t+ \frac{\dot{r}}{h}\, \hat{e}_r + r\dot{\phi}\, \hat{e}_\phi ,
\end{eqnarray}
where $h$ is defined by,
\begin{equation}
	\label{Eq6.17}
	h= \sqrt{\frac{r-1}{r}}.
\end{equation}
Employing \eqref{Eq6.13},  \eqref{Eq6.15}, and  \eqref{Eq6.5}, one would get,
\begin{eqnarray}
	\label{Eq6.18}
	\dot{\vec{\gamma}} 
	&=& \gamma \sqrt{\frac{r}{r-1}} \hat{e}_t - \sqrt{\frac{\gamma^2 r^3 - (r-1)(r^2+l^2)}{r^2(r-1)}} \hat{e}_r + \frac{l}{r} \hat{e}_\phi.
\end{eqnarray}
It also holds that $\dot{\vec{\gamma}}^2=-1$. We equate $\hat{e}_0$ to $ \dot{\vec{\gamma}}$, i.e.,
\begin{equation}
	\label{Eq6.19}
	\hat{e}_0 =   \gamma \sqrt{\frac{r}{r-1}} \hat{e}_t - \sqrt{\frac{\gamma^2 r^3 - (r-1)(r^2+l^2)}{r^2(r-1)}} \hat{e}_r + \frac{l}{r} \hat{e}_\phi.
\end{equation}
Let $\hat{e}_0$  be written as, 
\begin{equation}
	\label{Eq6.20}
	\hat{e}_0 =  \cosh{\alpha}\, \hat{e}_t -\sinh{\alpha} \,(\hat{e}_r \cos \beta+ \hat{e}_\phi \sin \beta),
\end{equation}
where,
\begin{subequations}
\label{Eq6.21}
	\begin{eqnarray}
		\cosh \alpha &=& \gamma \sqrt{\frac{r}{r-1}},\\ 
		\sinh \alpha &=& \sqrt{\frac{\gamma^2 r}{r-1}-1}, 
	\end{eqnarray}
\end{subequations}
and, 
\begin{subequations}
\label{Eq6.22}
	\begin{eqnarray} 		
		\cos \beta &=& \frac{1}{r}\sqrt{\frac{\gamma^2 r^3 - (r-1)(r^2+l^2)}{\gamma^2 r -r+1}},\\
		\label{Eq6.22b}	
		\sin \beta &=&\frac{l }{r}\sqrt{\frac{r-1}{\gamma^2  r -r + 1}}.
	\end{eqnarray}	
\end{subequations}
The succinct form of \eqref{Eq6.20} then allows us to write: 
\begin{equation}
	\label{Eq6.23}
	\hat{e}_1 = -\sinh\alpha \,\hat{e}_t +\cosh\alpha  \,(\hat{e}_r \cos \beta+ \hat{e}_\phi \sin \beta),
\end{equation} 
which holds that, 
\begin{eqnarray}
	\label{Eq6.24}
	\hat{e}_1\cdot\hat{e}_0 &=& 0,\\
	\label{Eq6.25}
	\hat{e}_1\cdot\hat{e}_1 & =& 1.
\end{eqnarray}  
The unit norm vector of $\hat{e}_2$ should be in the $\hat{e}_{r}-\hat{e}_{\phi}$-plane, and is perpendicular to $\hat{e}_0$ and $\hat{e}_1$. So it is given by, 
\begin{equation}
	\label{Eq6.26}
	\hat{e}_{2} = -\hat{e}_r \sin \beta+ \hat{e}_\phi \cos \beta,
\end{equation}
which holds $\hat{e}_{2}^2=1$. The unit norm vector $\hat{e}_3$ is given by,
\begin{eqnarray}
	\label{Eq6.27}
	\hat{e}_3&=& \hat{e}_\theta.
\end{eqnarray} 
An infinitesimal displacement $\delta\vec{x}$ in the standard spherical coordinates for $\theta=\frac{\pi}{2}$ is given by:
\begin{eqnarray}
	\label{Eq6.28}
	d\vec{x} =  h  dt \,\hat{e}_t+ \frac{1}{h} dr\, \hat{e}_r + r d\phi \,\hat{e}_\phi + r d\theta \,\hat{e}_\theta\,,
\end{eqnarray}
where $\hat{e}_t$, $\hat{e}_r$, $\hat{e}_\theta$ and $\hat{e}_\phi$ are the normal unit vectors. 
The infinitesimal displacement in Fermi coordinates is,
\begin{eqnarray}
	\label{Eq6.29}
	d\vec{x} =   dx^0 \hat{e}_0 + dx^a \hat{e}_a   \,,
\end{eqnarray}
where $\hat{e}_0$ and $\hat{e}_a$ are the unit vectors in  Fermi coordinates and  $\hat{e}_0$ is tangent to the geodesic $\gamma$. 
Utilizing \eqref{Eq6.19}, \eqref{Eq6.23}, \eqref{Eq6.26}, and \eqref{Eq6.27} in \eqref{Eq6.29} yields,  
\begin{eqnarray}
	\label{Eq6.30}
	d\vec{x} &=& ( \cosh\alpha\, dx^0- \sinh \alpha\, dx^1) \hat{e}_t \nonumber \\
	&+& (-\sinh \alpha \cos \beta\, dx^0 + \cosh \alpha \cos \beta \,dx^0- \sin \beta \,dx^2 )\hat{e}_r\nonumber\\
	&+& (-\sinh \alpha  \sin \beta\, dx^0 + \cosh \alpha \sin \beta\,  dx^1 + \cos \beta\, dx^2)\hat{e}_\phi\nonumber\\
	&+& \hat{e}_\theta \,dx^3.
\end{eqnarray}
We utilize \eqref{Eq6.21} to express $h$, which is defined in \eqref{Eq6.17}, i.e., 
\begin{equation}
	\label{Eq6.31}
	h = \frac{\gamma}{\cosh \alpha}.
\end{equation}
This allows us to rewrite \eqref{Eq6.28} as,
\begin{eqnarray}
	\label{Eq6.32}
	d\vec{x} =   \frac{\gamma}{\cosh \alpha} dt  \,\hat{e}_t+ \frac{\cosh \alpha}{\gamma} dr \,\hat{e}_r + r d\phi \,\hat{e}_\phi + r d\theta \,\hat{e}_\theta  \,,
\end{eqnarray}
that we  equate to \eqref{Eq6.30} to infer that,
\begin{eqnarray}
	\label{Eq6.33}
	dt & =&\frac{1}{\gamma}\cosh^2\alpha\, dx^0-\frac{1}{\gamma} \cosh \alpha \sinh \alpha\, dx^1,\\
	\label{Eq6.34}
	dr  &=&-\gamma \tanh \alpha \cos \beta\, dx^0 +\gamma \cos \beta \,dx^1-\gamma\frac{\sin \beta}{\cosh \alpha} \,dx^2 ,\\
	\label{Eq6.35}
	d\phi & =&-\frac{\sinh \alpha  \sin \beta}{r}\, dx^0 + \frac{\cosh \alpha \sin \beta}{r}\,  dx^1 + \frac{\cos \beta}{r}\, dx^2 , \\
	\label{Eq6.36}
	d\theta & = &\frac{dx^3}{r} .
\end{eqnarray} 
Their anti-symmetric wedge products are given by,
\begin{eqnarray}
	\label{Eq6.37}
	dt\wedge dr &=& \cos \beta  \,dx^0\wedge dx^1 - \sin \beta \, \cosh \alpha\, dx^0 \wedge dx^2 \nonumber \\&+& \sin \beta\,\sinh\alpha \,dx^1 \wedge dx^2,\\
	\label{Eq6.38}
	dt\wedge d\phi &=& \frac{\cosh\alpha\,\sin\beta}{r\gamma} dx^0 \wedge dx^1 
	+ \frac{\cosh^2 \alpha\,\cos\beta }{r\gamma}dx^0 \wedge dx^2\nonumber \\&-& \frac{\cosh \alpha \sinh\alpha \cos\beta}{r\gamma} dx^1 \wedge dx^2,\\
	\label{Eq6.39}
	dt \wedge d\theta &=& \frac{\cosh^2\alpha}{r\gamma} dx^0\wedge dx^3-\frac{\cosh \alpha \sinh \alpha}{r\gamma} \, dx^1\wedge dx^3,\\
	\label{Eq6.40}
	dr\wedge d\phi &=&  -\frac{\gamma \tanh \alpha}{r} dx^0 \wedge dx^2 + \frac{\gamma}{r} dx^1 \wedge dx^2,\\
	\label{Eq6.41}
	dr\wedge d\theta &=& -\frac{\gamma \tanh \alpha \cos \beta}{r} dx^0\wedge dx^3 +\frac{\gamma \cos \beta}{r}dx^1\wedge dx^3\nonumber\\&-&\frac{\gamma\sin \beta}{r\cosh \alpha} \,dx^2\wedge dx^3,\\
	\label{Eq6.42}
	d\phi\wedge d\theta &=& -\frac{\sinh \alpha  \sin \beta}{r^2} dx^0\wedge dx^3 + \frac{\cosh \alpha \sin \beta}{r^2}  dx^1\wedge dx^3 \nonumber\\&+& \frac{\cos \beta}{r^2} dx^2\wedge dx^3.
\end{eqnarray}
The above anti-symmetric wedge products can be substituted in \eqref{Eq6.9} to obtain the non-zero components of the Riemann tensor in Fermi coordinates,
%
\begin{subequations}
\label{Eq6.43}
	\begin{eqnarray}	
		R_{0101} &=& -\frac{1+ 3 \cos 2\beta}{4r^3},\\
		R_{0202} & =&\frac{1+ 3 \cos 2\beta}{4r^3}-\frac{3 l^2 +r^2}{2r^5} ,\\ 
		R_{0303} & =& \frac{3 l^2 +r^2}{2r^5},\\
		R_{0102} &=& \frac{3 \cosh \alpha\,\sin 2\beta}{2r^3}, 
	\end{eqnarray}
\end{subequations}
where $\alpha$ and $\beta$ are defined in \eqref{Eq6.21} and \eqref{Eq6.22}.
As a consistency check, we notice that the parity in the $x^3$ direction causes $R_{0103}=R_{0203}=0$.  As another consistency check, we note that  the $00$--component of the Ricci tensor evaluated on the geodesic, \eqref{Eq4.7} for $l=m=0$, vanishes too. It is advised to use Mathematica in simplification of what leads to \eqref{Eq6.43}. 

\section{Hydrogen atom in the Schwarzschild space-time geometry}
\label{Section7}
We would like to study the free fall of a hydrogen atom in its ground state on various geodesics in the Schwarzschild space-time geometry. We choose Fermi coordinates to describe the physics near the hydrogen atom. Ignoring $V_{{eff}}$, the Schr\"odinger equation for the electron bound to the hydrogen atom is given by,
\begin{equation}
	\label{Eq7.1}
	\left(-\frac{\hbar^2}{2m}\nabla^2 - \frac{e^2}{4\pi \epsilon_0 x}\right)\Psi^{(0)} = i\hbar \partial_0  \Psi^{(0)}\,,
\end{equation} 
where, 
\begin{equation}
	\label{Eq7.2}
	x^2 = (x_1)^2 + (x_2)^2 + (x_3)^2. 
\end{equation}
The normalized position wave functions, given in spherical coordinates, 
	\begin{eqnarray}
		\label{Eq7.3}
		x^1 &=& x \cos \theta,\\ \nonumber
		x^2 &=& x \sin \theta \cos\phi,\\ \nonumber
		x^3 &=& x \sin \theta \sin \phi,  
	\end{eqnarray} 
are, 
%
\begin{eqnarray}
	\label{Eq7.4}
	\Psi^{(0)}_{n,\ell,m}(x, \theta, \phi)&=& \sqrt{\left(\frac{2}{na_0}\right)^3 \frac{(n-\ell-1)!}{2n(n+\ell)!}} \nonumber\\ &\times& e^{-\rho/2} \rho^\ell L^{2\ell+1}_{n-\ell-1}(\rho) Y^m_\ell(\theta,\phi) e^{-\frac{i E_n \tau}{\hbar}},
\end{eqnarray}
where $\theta$ and $\phi$ are angles of the standard spherical coordinates in the lab, not to be taken as the same as what defined in \eqref{Eq6.2},  $x^0=\tau$ is the proper time, and, 
\begin{itemize}
	\item $E_n$ is the energy of the state $n$ and is given by, 
	\begin{equation}
		\label{Eq7.5}
		E_n= -\frac{m e^4}{2(4\pi \epsilon_0)^2 \hbar^2}\frac{1}{n^2},
	\end{equation}
	\item $\rho= \frac{2x}{n a_0}$,
	\item $a_0$ is the reduced Bohr radius,
	\begin{equation}
		\label{Eq7.6}
		a_0= \frac{4\pi \epsilon_0 \hbar^2}{m e^2}=0.53~{\AA},
	\end{equation} 
	\item $L^{2\ell+1}_{n-\ell-1}(\rho)$ is a generalized Laguerre polynomial of degree $n-\ell-1$,
	\item $Y^{m}_\ell$ is a spherical harmonic function of degree $\ell$ and order $m$. 
\end{itemize}
It is convenient to utilize the Dirac notation and write,
\begin{equation}
	\label{Eq7.7}
	\Psi^{(0)}_{n,\ell,m}(x, \theta, \phi) = e^{-\frac{i E_n \tau}{\hbar}} \braket{x}{n,\ell,m}.
\end{equation}
Note that the generalized Laguerre polynomials are defined differently by different authors. The usage here is consistent with the definitions used by Mathematica. The quantum numbers can take the following values, 
\begin{eqnarray}
	\label{Eq7.8}
	n&=& 1, 2, 3, \ldots \\
	\label{Eq7.9}
	\ell&=& 0,1,2, \ldots, n-1,\\
	\label{Eq7.10}
	m &=& -\ell, \ldots, +\ell\,,
\end{eqnarray}
where $n$ is the principle quantum number, $\ell$ is the azimuthal quantum number, and $m$ is named the magnetic quantum number.  The ground state is known to be,
\begin{equation}
	\label{Eq7.11}
	\Psi^{(0)}_{1,0,0}=e^{- \frac{i E_1 \tau}{\hbar}} \braket{x}{1,0,0}= \frac{ e^{- \frac{i E_1 \tau}{\hbar}} }{\sqrt{\pi} {a_0}^{\frac{3}{2}}} e^{- \frac{x}{a_0}},
\end{equation}
where $E_1$ is the energy of the ground state and is,
\begin{eqnarray}
	\label{Eq7.12}
	E_1  &=& -\frac{m e^4}{2 (4\pi \epsilon_0)^2 \hbar^2} = - 13.6\, eV.
\end{eqnarray}
The first excited states are, 
%
\begin{subequations}
	\begin{eqnarray}
			\label{Eq7.13}
		\Psi^{(0)}_{2,0,0} &=&e^{- \frac{i E_2 \tau}{\hbar}} \braket{x}{2,0,0}=   \frac{ e^{- \frac{i E_2 \tau}{\hbar}} }{4\sqrt{2\pi} {a_0}^{\frac{3}{2}}}\left(2- 	\frac{x}{a_0}\right) e^{-\frac{x}{2 a_0}},\\ 
		\Psi^{(0)}_{2,1,-1} &=&e^{- \frac{i E_2 \tau}{\hbar}} \braket{x}{2,1,-1}= \frac{ e^{- \frac{i E_2 \tau}{\hbar}} }{4\sqrt{\pi} {a_0}^{\frac{5}{2}}}
		x e^{-\frac{x}{2a_0}-i\phi } \sin\theta,\\ 
		\Psi^{(0)}_{2,1,0} &=&e^{- \frac{i E_2 \tau}{\hbar}} \braket{x}{2,1,0}= \frac{ e^{- \frac{i E_2 \tau}{\hbar}} }{4\sqrt{2\pi} {a_0}^{\frac{5}{2}}}
		x e^{-\frac{x}{2a_0}} \cos\theta,\\ 
		\Psi^{(0)}_{2,1,1} &=&e^{- \frac{i E_2 \tau}{\hbar}} \braket{x}{2,1,1}=  \frac{ e^{- \frac{i E_2 \tau}{\hbar}} }{4\sqrt{\pi} {a_0}^{\frac{5}{2}}}
		x e^{-\frac{x}{2a_0}+i\phi } \sin\theta,
	\end{eqnarray}
\end{subequations}
%
where we have used the Dirac notation. Note that, in our notation, $\ket{n,\ell,m}$ represents the state of the electron at $\tau=0$.

We would like to calculate the effective potential that we obtained in section \ref{Section5} for a general time-like geodesic that we studied in section \ref{Section6.1}. In so doing, we notice that utilizing  \eqref{Eq5.23}, \eqref{Eq6.43}, and  \eqref{Eq7.3} gives,
\begin{eqnarray}
	\label{Eq7.14}
	V_{{eff}}
	= \frac{m x^2}{2 r^3}&& \left( -\frac{1+ 3 \cos 2\beta}{4} \cos^2 \theta+ \frac{3 l^2 +r^2}{2r^2} \sin^2\theta\sin^2\phi \right.\nonumber \\
	&&~~+ (\frac{1+ 3 \cos 2\beta}{4}-\frac{3 l^2 +r^2}{2r^2}) \sin^2 \theta \cos^2\phi\nonumber\\
	 &&~~~\left.+  \frac{3 \cosh \alpha\,\sin 2\beta}{2} \sin2\theta \cos\phi \right),
\end{eqnarray}
which can be simplified to, 
\begin{eqnarray}
	\label{Eq7.15}
	V_{{eff}}= -\frac{m c^2 r_s x^2}{32 r^3}&&\Big{(}(1+3 \cos 2\beta)(1+3 \cos 2\theta) +
	\nonumber\\ &&+6(\frac{4 l^2}{r^2}+1 - \cos2\beta)\cos 2\phi \sin^2\theta\nonumber\\
	&&- 24  \cosh\alpha \sin 2\beta \cos\phi \sin 2 \theta \Big{)}, 
\end{eqnarray}
where $m$ is the mass of particle; $c$ and the Schwarzschild radius ($r_s$) are recovered;  $x$ is the distance from centre of lab (the time-like geodesic);  $r$ is distance of the lab from centre of the Schwarzschild geometry; $l$ is the constant of the time-like geodesic defined in  \eqref{Eq6.5}; $\alpha$ and $\beta$ are given in \eqref{Eq6.21} and \eqref{Eq6.22}; and $\theta$ and $\phi$ are the standard spherical angles in the lab defined in \eqref{Eq7.3}.

\section{Absorption by  a black hole: a radially falling hydrogen atom}
\label{Section8}
We first consider a hydrogen atom at $r=a$ in its ground state radially falling down, and we compute what would be the probability of transition to higher modes (excited states) as the hydrogen atom falls down inside the Schwarzschild black hole.  We notice that the radial geodesic is described by $l=0$, and consequently, $\beta=0$. The effective gravitational potential for a radially falling hydrogen atom, therefore, is simplified to, 
\begin{eqnarray}
	\label{Eq8.1}
	V_{{eff}}= -\frac{m c^2 r_s x^2}{8 r^3}\left(1+3 \cos 2\theta\right). 
\end{eqnarray}
Equation \eqref{Eq6.15} is simplified to, 
\begin{equation}
	\label{Eq8.2}
	\dot{r}= -c\,\sqrt{\gamma^2 -1 + \frac{r_s}{r} },
\end{equation}
where $\gamma$ is the Lorentz factor for the hydrogen atom in the asymptotic infinity, as defined in \eqref{Eq6.12}, and $c$ and $r_s$ are recovered. We assume that the hydrogen atom in its ground state is at $\tau=0$ and is located at radius of $a$. At $\tau=\tau[r]$, it will be at $r$; thus, 
\begin{equation}
	\label{Eq8.3}
	\tau[r] = \int_a^r \frac{dr}{\dot{r}}= -\frac{1}{c}\int_a^r \frac{dr}{\sqrt{\gamma^2-1 + \frac{r_s}{r}}}.
\end{equation}
Let us first study the indefinite integral of, 
\begin{equation}
	\label{Eq8.4}
	\tau[r] = \int \frac{dr}{\dot{r}}= -\frac{1}{c}\int \frac{dr}{\sqrt{\gamma^2-1 + \frac{r_s}{r}}}
\end{equation}
assuming that $\gamma\neq1$\footnote{
	We will consider the case of $\gamma=1$ in \eqref{Eq8.37}. } and
defining, 
\begin{equation}
	\label{Eq8.5}
	r= \tilde{r} \frac{r_s}{\gamma^2-1}.
\end{equation}
This change of variable allows us to represent the integration in \eqref{Eq8.4} by, 
\begin{equation}
	\label{Eq8.6}
	c\tau[r]=c\tau[\tilde{r}]= -\frac{r_s}{(\gamma^2-1)^{\frac{3}{2}}}\left(\sqrt{\tilde{r}(\tilde{r}+1)}+ \log\left(\sqrt{\tilde{r}+1}-\sqrt{\tilde{r}}\right)\right).
\end{equation}
Now let us consider a hydrogen atom which is in the state of $\ket{\alpha}$ at $r=a$ and $\tau=0$. As the hydrogen atom moves along the time-like geodesic of its proton, it can be excited to higher states due to the change of the curvature of the space-time geometry around the proton. The transition amplitude to the state $\ket{\beta}$ at radius $b$, given in \eqref{Eq5.27}, can be rewritten as,
\begin{equation}
	\label{Eq8.7}
	A_{\alpha\beta}= -\frac{i}{\hbar} \int_{a}^b \frac{dr}{\dot{r}} \bra{\beta}V_{{eff}}\ket{\alpha} 	e^{-\frac{i \left(E_\alpha-E_\beta\right)}{c\hbar}(c\tau[r]-c\tau[a])}.
\end{equation}
We would like to study the transition from the ground state of the hydrogen atom. Therefore, we set,
\begin{equation}
	\label{Eq8.8}
	\ket{\alpha}=\ket{1,0,0}.
\end{equation}
For the transition to $\ket{\beta}=\ket{n, \ell, m}$, it yields, 
\begin{eqnarray}
	\label{Eq8.9}
	\bra{n,\ell,m}V_{{eff}}\ket{1,0,0} &=& -\frac{m c^2 r_s}{8 r^3}\bra{n,\ell,m}x^2(1+3 \cos 2\theta)\ket{1,0,0}.
\end{eqnarray}
Noticing that $(1+3 \cos 2\theta)$ is proportional to the second Legendre polynomials of $\cos \theta$, and considering the properties of the states of the hydrogen atoms, it yields,
\begin{eqnarray}
	\label{Eq8.10}
	\bra{n,\ell,m}V_{{eff}}\ket{1,0,0} &=& -\frac{m c^2 r_s}{8 r^3}\delta^{m0}\delta^{\ell2}\bra{n,2,0}x^2(1+3 \cos 2\theta)\ket{1,0,0},
\end{eqnarray} 
where $\delta^{m0}$ and $\delta^{\ell2}$ represent the Kronecker delta function. The same mechanism demands the following rules for non-zero transition from $\ket{n,\ell,m}$ to $\ket{n',\ell',m'}$:
%
	\begin{eqnarray}
		\label{Eq8.11}
		m&=&m',\\
		\ell-\ell'&=&\pm2.
	\end{eqnarray}
%
The first allowed transition from the ground state, therefore, is to $\ket{3,2,0}$, which yields, 
%
\begin{equation}
	\label{Eq8.12}
	\bra{3,2,0}V_{{eff}}\ket{1,0,0} =-\frac{81\sqrt{3} m c^2 r_s}{256\sqrt{2} r^3} (a_0)^2,
\end{equation} 
where $a_0$ is the reduced Bohr radius defined in \eqref{Eq7.6}. Here it is observed that the dominant transition is to the 3d state. The transition to the 3d state is observed in a different context\cite{Hu:2021sfb}. 

Utilizing \eqref{Eq8.2}, \eqref{Eq8.12} in \eqref{Eq8.7} yields, 
\begin{equation}
	\label{Eq8.13}
	A_{\ket{1,0,0}\to\ket{3,2,0}}=  -\frac{81i\sqrt{3} m c r_s (a_0)^2}{256\sqrt{2}\hbar}  e^{-\frac{i \Delta E}{c\hbar}c\tau[a]} \int_{a}^b \frac{dr}{r^3}  	\frac{e^{-\frac{i \Delta E}{c\hbar}c\tau[r]}}{\sqrt{\gamma^2-1+\frac{r_s}{r}}},
\end{equation}
where $\Delta E$ is the energy difference between $\ket{1,0,0}$ and $\ket{3,2,0}$,
\begin{equation}
	\label{Eq8.14}
	\Delta E = E_1 - E_3= -12.09\,eV.
\end{equation}
Expressing \eqref{Eq8.13} in terms of $\tilde{r}$ (defined in \eqref{Eq8.5}) yields, 
%
\begin{equation}
	\label{Eq8.15}
	\left|A_{\ket{1,0,0}\to\ket{3,2,0}}\right|=  \frac{81\sqrt{3} m c (a_0)^2}{256\sqrt{2}\hbar} \frac{(\gamma^2-1)^{\frac{3}{2}}}{r_s}  \left|\int_{\tilde{a}}^{\tilde{b}} d\tilde{r}\frac{e^{-\frac{i \Delta E}{c\hbar}c\tau[\tilde{r}]}}{\sqrt{\tilde{r}^3(1+\tilde{r})}}\right| ,
\end{equation}
where, 
	\begin{eqnarray}
			\label{Eq8.16}
		\tilde{a}&=& \frac{\gamma^2-1}{r_s} a,\\\nonumber
		\tilde{b}&=& \frac{\gamma^2-1}{r_s} b.
	\end{eqnarray} 
%
Note that it is assumed that $\gamma^2-1\neq 0$. For $\gamma=1$, \eqref{Eq8.5} cannot be used.
\begin{figure}[t]
	\begin{center}
		\includegraphics[width=0.8\textwidth]{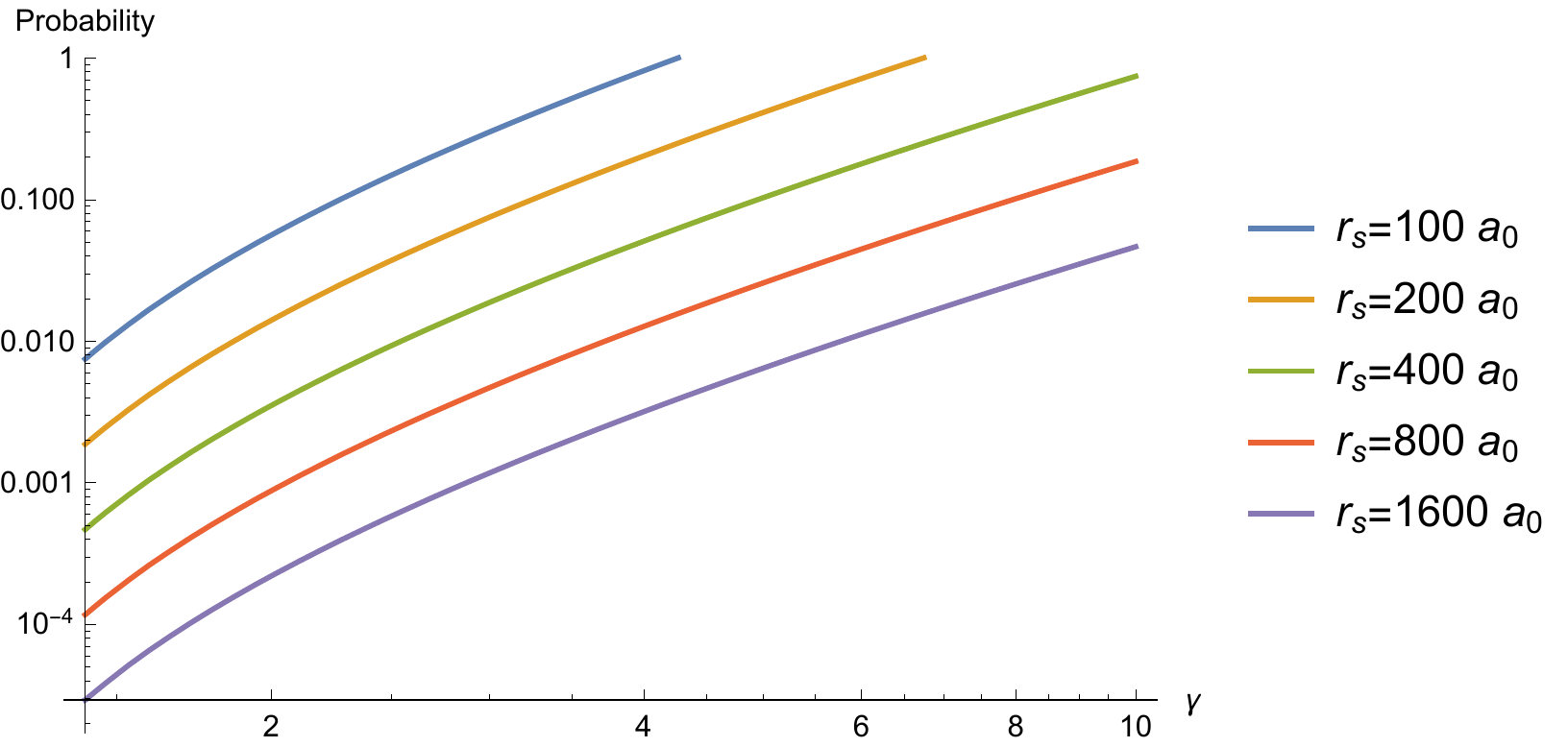}
	\end{center}
	\caption{Probability of transition of a hydrogen atom from its ground state at the asymptotic infinity to $\ket{3,2,0}$ on the event horizon as the hydrogen atom radially falls into the black hole versus the Lorentz factor of the hydrogen at the asymptotic infinity. Both axes are in logarithmic scale, and $\gamma\in[\sqrt{2},10]$.}
	\label{fig:fig1}
\end{figure}

Let us calculate the numerical values of the coefficients present in \eqref{Eq8.15}. It holds that\footnote{Click \href{https://www.wolframalpha.com/input/?i=81*sqrt(3)*(mass of electron)*(light speed)*(bohr radius)^2/(256*sqrt(2)*hbar)}{here} to have WolframAlpha perform the numerical substitution}, 
\begin{equation}
	\label{Eq8.17}
	\frac{81\sqrt{3} m c  (a_0)^2}{256\sqrt{2} \hbar} = 53.10 a_0, 
\end{equation}
and\footnote{Click \href{https://www.wolframalpha.com/input/?i=(13.6 -13.6 /9)eV/(hbar*c)*(Bohr radius)}{here} to have the numerical value by WolframAlpha}, 
\begin{equation}
	\label{Eq8.18}
	\frac{\Delta E}{c\hbar} = 3.24 \times 10^{-3} \frac{1}{a_0}=\frac{1}{308a_0}.
\end{equation}
Equation \eqref{Eq8.13} can be expressed as,  
\begin{equation}
	\label{Eq8.19}
	\left|A_{\ket{1,0,0}\to \ket{3,2,0}}\right|=  53.10 \frac{(\gamma^2-1)^{\frac{3}{2}}a_0}{r_s}  \left|\int_{\tilde{a}}^{\tilde{b}} d\tilde{r} \frac{e^{-i \frac{c\tau[\tilde{r}]}{308a_0} }}{\sqrt{\tilde{r}^3(1+\tilde{r})}}\right|.
\end{equation}
We would like to calculate the transition amplitude for a hydrogen atom at the asymptotic infinity which falls inside the event horizon. So we should set $a=\infty$ and $b=r_s$. Equation \eqref{Eq8.16} then gives $\tilde{a}=\infty$ and $\tilde{b}=\gamma^2-1$. The amplitude then follows,
\begin{equation}
	\label{Eq8.20}
	\left|A_{\ket{1,0,0}\to \ket{3,2,0}}\right|=  53.10 \frac{(\gamma^2-1)^{\frac{3}{2}}a_0}{r_s}  \left|\int_{\gamma^2-1}^{\infty} dx \frac{e^{-if(x) }}{\sqrt{x^3(1+x)}}\right|,
\end{equation}
where, 
\begin{eqnarray}
	\label{Eq8.21}
	f(x)&=& - A g(x),\\
	\label{Eq8.22}
	A &=& \frac{0.00324}{(\gamma^2-1)^{\frac{3}{2}}}\frac{r_s}{a_0},\\
	\label{Eq8.23}
	g(x)&=&\sqrt{x(x+1)}+ \log\left(\sqrt{x+1}-\sqrt{x}\right).
\end{eqnarray}
Due to the factor of $\frac{a_0}{r_s}$ in the front of the integral, and the factor of $\frac{r_s}{a_0}$ in the power of the exponential factor in the integral, the amplitude is very negligible unless $r_s$ is not much larger that $a_0$.

Figure \ref{fig:fig1} depicts the transition amplitude for $r_s=100 a_0$ to $r_s=1600 a_0$ and the Lorentz factor $\gamma=\sqrt{2}$ to $\gamma=10$. Let it be highlighted that, for $\gamma$ very close to unity, the error in the numerical calculation rises; therefore, we choose $\gamma\ge\sqrt{2}$. 

\subsection{The ultra-relativistic regime}
\label{Section8.1}
The large $\gamma$ limit in \eqref{Eq8.20} corresponds to the large $x$ limit of $g(x)$ in \eqref{Eq8.23}:
\begin{equation}
	\label{Eq8.24}
	g(x)= x - \frac{1}{2} \log x +\frac{1}{2}(1-2 \log 2)+ O\left(\frac{1}{x}\right).
\end{equation}
For large $x$, the factor of $(\sqrt{x^3(1+x)})^{-1}$ in the integral of  \eqref{Eq8.20} yields, 
\begin{equation}
	\label{Eq8.25}
	\frac{1}{\sqrt{x^3(1+x)}}=\frac{1}{x^2} \left(1+ O\left(\frac{1}{x}\right)\right).
\end{equation}
Utilizing \eqref{Eq8.24} and \eqref{Eq8.25} yields, 
\begin{equation}
	\label{Eq8.26}
	\left|\int_{\gamma^2-1}^{\infty} dx \frac{e^{-if(x) }}{\sqrt{x^3(1+x)}}\right| = \left|\int_{\gamma^2-1}^{\infty} \frac{dx}{x^2} e^{i (x-\frac{1}{2}\log x)}\left(1+ O(\frac{1}{x})\right) \right|.
\end{equation}
Ignoring $O(\frac{1}{x})$ in \eqref{Eq8.26} leads to an error which is less than $\frac{1}{\gamma^2-1}$. For $\gamma>10$, the error will be less than one percent. We assume that $\gamma$ is greater than $10$ and ignore $O(\frac{1}{x})$ in \eqref{Eq8.26}. This enables us to calculate \eqref{Eq8.26} with a precision of one percent:
\begin{eqnarray}
	\label{Eq8.27}
	\left|\int_{\gamma^2-1}^{\infty} dx \frac{e^{-if(x) }}{\sqrt{x^3(1+x)}}\right| &=& \left|\int_{\gamma^2-1}^{\infty} \frac{dx}{x^2} e^{i A(x-\frac{1}{2}\log x) } \right|\nonumber\\
	&=&\frac{1}{\gamma^2-1} \left| {E}\left(2 + \frac{i A}{2},- i (\gamma^2-1) A\right) \right|,
\end{eqnarray}
where ${E}(n,z)$ is the exponential integral function, i.e.,
\begin{equation}
	\label{Eq8.28}
	{E}(n,z)= \int_1^\infty \frac{e^{-zt}}{t^n} dt.
\end{equation}
Utilizing \eqref{Eq8.27} in \eqref{Eq8.20} yields, 
\begin{equation}
	\label{Eq8.29}
	\left|A_{\ket{1,0,0}\to \ket{3,2,0}}\right|_\gamma=  53.10 \frac{(\gamma^2-1)^{\frac{1}{2}}a_0}{r_s}  \left|{E}\left(2 + \frac{i A}{2},- i (\gamma^2-1) A\right)\right|.
\end{equation}
For a large $\gamma$, 
\begin{eqnarray}
	\label{Eq8.30}
	A &=&  \frac{0.00324}{\gamma^3} \frac{r_s}{a_0},\\
	\label{Eq8.31}
	(\gamma^2-1) A &=& \frac{0.00324}{\gamma} \frac{r_s}{a_0}.
\end{eqnarray}
In the limit of a vanishing $A$, \eqref{Eq8.29} is given by,
\begin{equation}
	\label{{Eq8.32}}
	\left|A_{\ket{1,0,0}\to \ket{3,2,0}}\right|_\gamma=  53.10 \frac{\gamma a_0}{r_s}  \left|{E}\left(2,- i  \frac{0.00324}{\gamma} \frac{r_s}{a_0}\right)\right|.
\end{equation}
When $\gamma$ is large such that it holds that,
\begin{equation}
	\label{Eq8.33}
	0.00324\frac{r_s}{a_0}\ll  \gamma,
\end{equation}
then \eqref{{Eq8.32}} can be approximated to,
\begin{equation}
	\label{Eq8.34}
	\left|A_{\ket{1,0,0}\to \ket{3,2,0}}\right|_\gamma=  53.10  \frac{\gamma a_0}{r_s}.
\end{equation}
This means that there exists the critical Lorentz factor of, 
\begin{equation}
	\label{Eq8.35}
	\gamma_c= \frac{r_s}{53.10 a_0},
\end{equation}
such that, for all Lorentz factors larger than $\gamma_c$, the perturbative approach predicts that the hydrogen atom gets excited from its ground state as it falls into the black hole. However, let it be emphasized that, as $\gamma$ approaches $\gamma_c$ from a smaller value, the perturbation breaks down and a non-perturbative approach will be needed to calculate the exact value of probability of excitation.

\subsection{The classical regime}
\label{Section8.2}
\begin{figure}[t]
	\begin{center}
		\includegraphics{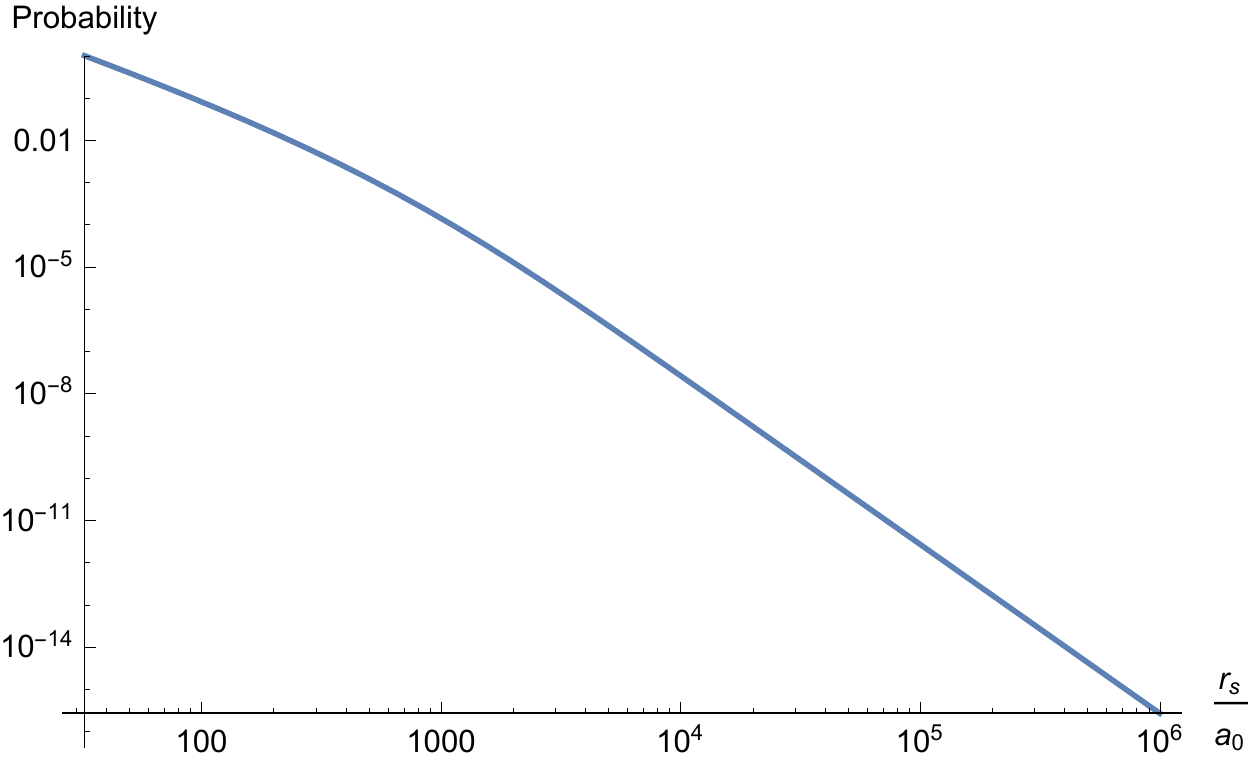}
	\end{center}
	\caption{Probability of transition of a  hydrogen atom from its ground state at the asymptotic infinity to $\ket{3,2,0}$ on the event horizon as the hydrogen atom radially falls into a black hole with zero velocity at infinity. The horizontal line is the radius of event horizon divided by $a_0$, where $a_0$ is the Bohr radius.}
	\label{fig:fig2}
\end{figure}
For the case of $\gamma=1$, Eq. \eqref{Eq8.4} can be simplified to,
\begin{equation}
	\label{Eq8.36}
	c\tau[r] = -\frac{2 r^{\frac{3}{2}}}{3 r_s^{\frac{1}{2}}},
\end{equation}
and \eqref{Eq8.13} is given by,
\begin{eqnarray}
	\label{Eq8.37}
	\left|A_{\ket{1,0,0}\to\ket{3,2,0}}\right|_{\gamma=1}=  \frac{81i\sqrt{3} m c\,r_s (a_0)^2}{256\sqrt{2}\hbar}  \frac{e^{-\frac{i \Delta E}{c\hbar}c\tau[a]}}{\sqrt{r_s}} \int_{a}^b \frac{dr}{r^{\frac{5}{2}}}  \exp\left(\frac{2i \Delta E}{3c\hbar\sqrt{r_s}}r^{\frac{3}{2}}\right).\nonumber\\
\end{eqnarray}
Therefore, the amplitude of the transition from the ground state of the hydrogen atom at the asymptotic infinity to the excited state of $\ket{3,2,0}$ on the event horizon, for the case of $\gamma=1$, is given by, 
\begin{equation}
	\label{Eq8.38}
	\left|A_{\ket{1,0,0}\to\ket{3,2,0}}\right|_{\gamma=1}= 53.10\, a_0\sqrt{r_s} \left|\int_{r_s}^\infty \frac{dr}{r^{\frac{5}{2}}}  	\exp\left(i\frac{0.00216}{a_0\sqrt{r_s}}r^{\frac{3}{2}}\right)\right|,
\end{equation}
where \eqref{Eq8.17} and \eqref{Eq8.18} are utilized. Let us set,
\begin{eqnarray}
	\label{Eq8.39}
	r &=& r_s u, \\
	\label{Eq8.40}
	u_s &=& 0.00216\, \frac{r_s}{a_0}.
\end{eqnarray}
Therefore, \eqref{Eq8.38} can be rewritten as,
\begin{equation}
	\label{Eq8.41}
	\left|A_{\ket{1,0,0}\to\ket{3,2,0}}\right|_{\gamma=1}=  \frac{0.1147}{u_s} 	\left|\int_{1}^\infty \frac{du}{u^{\frac{5}{2}}}  	\exp\left(iu_s u^{\frac{3}{2}}\right)	\right|.
\end{equation}
The integration in \eqref{Eq8.41} can be expressed in terms of the incomplete $\Gamma$ function\footnote{Incomplete $\Gamma$ function is defined by 
	\begin{displaymath}
		\Gamma(a,z)=\int_z^\infty t^{a-1} e^{-t} dt.
	\end{displaymath}
	The $\Gamma$ function can be evaluated to an  arbitrary precision by Wolfram Mathematica.}, 
\begin{equation}
	\label{Eq8.42}
	\left|A_{\ket{1,0,0}\to\ket{3,2,0}}\right|_{\gamma=1}=  \frac{0.0765}{u_s} 	\left|e^{i u_s}+ i u_s \Gamma(0, -i u_s)	\right|.
\end{equation}
The probability is smaller than $10\%$ for $r_s>91.44\,a_0$, which exceeds unity for $r_s<32.512\, a_0$. We cannot trust this perturbative computation when the probability reaches $1$. Therefore, we set $r_s>100\, a_0$. We would like to study the asymptotic behaviour of \eqref{Eq8.42} for large $u_s$. We first notice that  due to \eqref{Eq8.40}, the large $u_s$ is mapped to,
\begin{equation}
	\label{Eq8.43}
	462.963 a_0\ll r_s.
\end{equation} 
The amplitude in Eq. \eqref{Eq8.42} then can be approximated to,
\begin{equation}
	\label{Eq8.44}
	\left|A_{\ket{1,0,0}\to\ket{3,2,0}}\right|_{\gamma=1~\&~462.963 a_0\ll r_s} = 16389.5 \left(\frac{ a_0}{r_s}\right)^2\approx \left(\frac{ 2^7 a_0}{r_s}\right)^2.
\end{equation}
Figure \ref{fig:fig2} depicts the probability of transition to $\ket{3,2,0}$ as a function of the Schwarzschild radius of the black hole when a stationary hydrogen atom at its ground state in the asymptotic infinity radially falls into the black hole.

\subsection{Relativistic enhancement factor}
\label{Section8.3}
\begin{figure}[t]
	\begin{center}
		\includegraphics[width=0.8\textwidth]{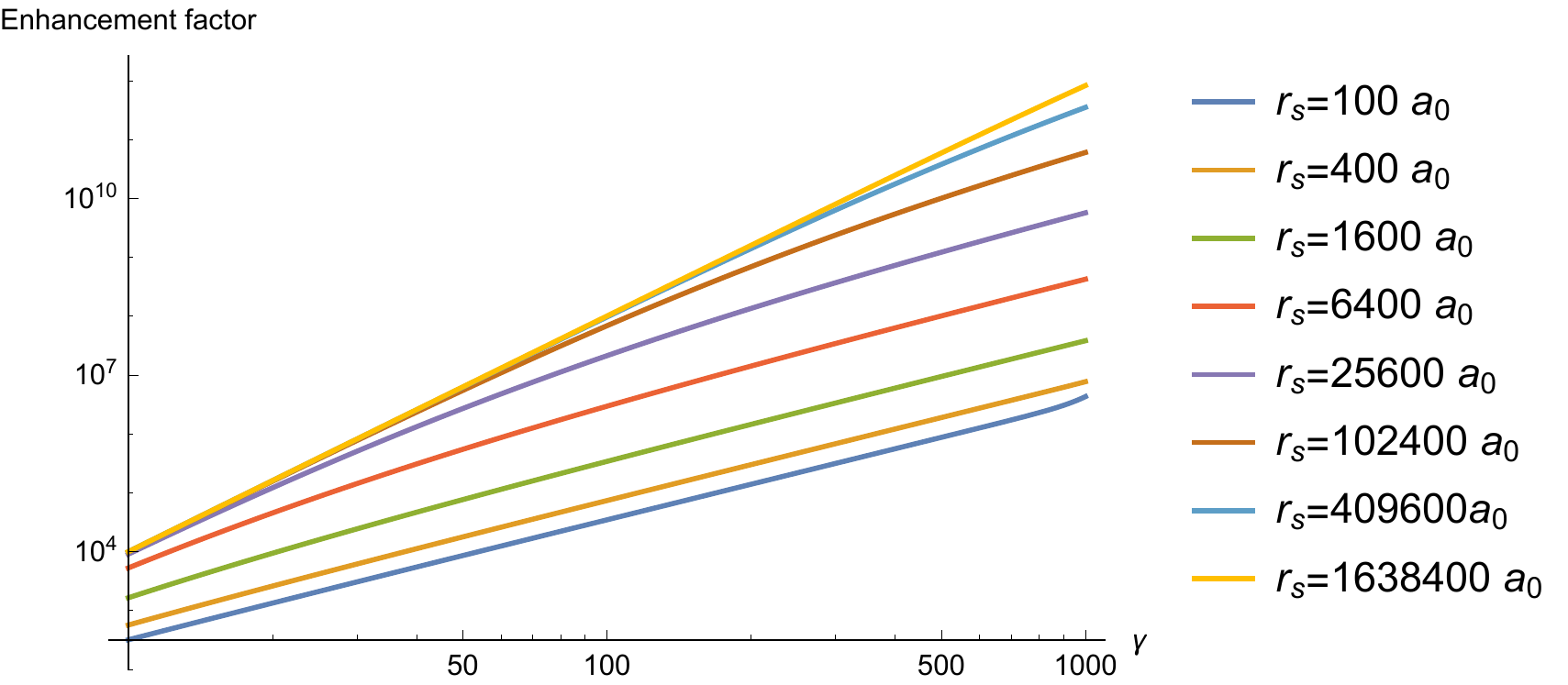}
	\end{center}
	\caption{The enhancement factor in terms of the Lorentz factor for a set of values of the Schwarzschild radius.}
	\label{fig:fig3}
\end{figure}
Equation \eqref{Eq8.29} is the amplitude of transition for an ultra-relativistic hydrogen atom with Lorentz factor $\gamma$ falling from asymptotic infinity radially into a black hole. Equation \eqref{Eq8.42} is the amplitude of transition for a stationary hydrogen atom at asymptotic infinity radially falling inside the black hole.  In order to better understand the effect of $\gamma$, let us define the relativistic enhancement factor by, 
\begin{equation}
	\label{Eq8.45}
	{Ef}(r_s,\gamma) = \left(\frac{\left|A_{\ket{1,0,0}\to \ket{3,2,0}}\right|_{\gamma}}{\left|A_{\ket{1,0,0}\to\ket{3,2,0}}\right|_{\gamma=1}}\right)^2
\end{equation}
The enhancement factor ${Ef}$ tells us how the probability of transition to the state of $\ket{3,2,0}$ is affected by increasing $\gamma$.  For large $r_s$ as defined by \eqref{Eq8.43}, and for large $\gamma$ as defined in \eqref{Eq8.33}, the enhancement factor simplifies to,

\begin{equation}
	\label{Eq8.46}
	{Ef}(r_s,\gamma)\approx  \left(\frac{\gamma r_s}{308  a_0}\right)^2.
\end{equation}
The enhancement factor for a set of general $\gamma$ and $r_s$ is shown in Fig. \ref{fig:fig3}. The transition probability of a stationary hydrogen atom falling into a black hole with the Schwarzschild radius $r_s=10^5 a_0$ is $2.69 \times 10^{-12}$.  The enhancement factor for $\gamma=1000$, however, increases this probability to $15.5\%$.

\section{A hydrogen atom deflected by a black hole}
\label{Section9}
We would like to consider a hydrogen atom deflected by the black hole. The hydrogen atom starts moving toward the black hole from radius of $r=\infty$ on a general time-like geodesic identified by $l$ and $\gamma$. So $\dot{r}$ is given by \eqref{Eq6.15}:
\begin{equation}
	\label{Eq9.1}
	\dot{r}= \pm  \sqrt{\gamma^2 -\frac{(r-r_s)(r^2+l^2)}{r^3} },
\end{equation}
where $r_s$ is recovered. $\dot{r}$ is negative until the hydrogen atom reaches a minimum distance from the black hole $r=r_{{min}}$, where $\dot{r}$ vanishes,
\begin{equation}
	\label{Eq9.2}
	\dot{r}|_{r=r_{min}}= -\sqrt{\gamma^2 -\frac{(r_{{min}}-r_s)(r_{{min}}^2+l^2)}{r_{{min}}^3} }=0,
\end{equation}
which can be solved to express $l$ in terms of $r_{min}$,
\begin{equation}
	\label{Eq9.3}
	l= r_{min} \sqrt{\frac{r_{min}}{r_{min}-r_s}\gamma^2-1}.
\end{equation}
It is easier to work in the length unit of, 
\begin{equation}
	\label{Eq9.4}
	r_{{min}}=1,
\end{equation}
where $l$ is given by,
%
\begin{equation}
	\label{Eq9.5}
	l= \sqrt{\frac{\gamma^2}{1-r_s}-1},
\end{equation}
where as the particle does not fall into the black hole,
\begin{equation}
	\label{Eq9.6}
	r_s<1,
\end{equation}	
is understood, which beside $1\leq \gamma$ guarantees that $l$ is a positive real number. Substituting the value of $l$ into $\dot{r}$ yields,
\begin{equation}
	\label{Eq9.7}
	\dot{r}=\pm  \sqrt{\gamma ^2 \left(1-\frac{r-r_s}{r^3 (1-r_s)}\right)-\frac{\left(r^2-1\right) (r-r_s)}{r^3}},
\end{equation}
where it is easy to observe that $\dot{r}|_{r=1}=0$ and $\dot{r}|_{r=+\infty}=\pm\gamma$. 

Very close to $r=1$, $r$ can be written as,
\begin{equation}
	\label{Eq9.8}
	r=1+\epsilon,
\end{equation}
where $\epsilon$ is a small number. The Taylor expansion of $\dot{r}$ given in \eqref{Eq9.7} for small $\epsilon$ yields, 
%
\begin{equation}
	\label{Eq9.9}
	\frac{1}{|\dot{r}|}=\kappa \epsilon^{-\frac{1}{2}}+ O(\epsilon^{\frac{1}{2}}),
\end{equation}
where,
\begin{equation}
	\label{Eq9.10}
	\kappa= \sqrt{\frac{1-r_s}{(2-3 r_s)\gamma^2 - 2 (1-r_s)^2}}.
\end{equation}
Since $\dot{r}$ is a real quantity, the following conditions must be fulfilled:
\begin{eqnarray}
	\label{Eq9.11}
	r_s &\leq &\frac{2}{3},\\
	\label{Eq9.12}
	\frac{\sqrt{2}(1-r_s)}{\sqrt{2-3 r_s}}&\leq &\gamma.
\end{eqnarray} 
When the minimum distance of the hydrogen atom to the centre of the black hole is larger than twice the Schwarzschild radius, or equivalently  due to the chosen length unit in \eqref{Eq9.4}:
%
\begin{subequations}
	\label{9.13series}
	\begin{eqnarray}	\label{Eq9.13}
		r_s \leq \frac{1}{2},
	\end{eqnarray}
then $\gamma$ can have any values. However there exists a lower bound on $\gamma$ if the hydrogen atom's distance to the centre of the black hole becomes smaller than twice of the Schwarzschild radius,
	\begin{eqnarray}
		\frac{1}{2}\leq r_s\leq \frac{2}{3},\\
		\frac{\sqrt{2}(1-r_s)}{\sqrt{2-3 r_s}}\leq \gamma.
	\end{eqnarray}
\end{subequations}
This means that a stationary atom at the asymptotic infinity falls into the black hole if its minimum distance to the black hole becomes smaller than twice the radius of the Schwarzschild black hole. 
%
We further notice that $r_s=\frac{2}{3}$ represents the photon sphere in the Schwarzschild geometry in the length unit chosen by \eqref{Eq9.4}. Interested readers may look at the literature for  further information on  the photon's sphere \cite{Claudel:2000yi}. For any finite values of $\gamma$, there exists a region of space around the black hole outside the photon's sphere that cannot be studied by scattering a hydrogen atom thrown toward the black hole. For any $r_s$ and $\gamma$  outside \eqref{9.13series}, the hydrogen atom falls inside the black hole instead of being deflected to the asymptotic infinity. 

For the length unit given by \eqref{Eq9.4}, $\alpha$ and $\beta$ defined in \eqref{Eq6.21} and \eqref{Eq6.22} are represented by, 
\begin{subequations}
	\begin{eqnarray}	\label{Eq9.14}
		\cosh \alpha &=& \gamma \sqrt{\frac{r}{r-r_s}},\\
		\sinh \alpha &=& \sqrt{\frac{\gamma^2 r}{r-r_s}-1}, 
	\end{eqnarray}
\end{subequations}
and, 
\begin{subequations}
	\begin{eqnarray}
		\label{Eq9.15}
		\cos \beta &=& \frac{1}{r}\sqrt{\frac{\gamma^2 r^3 - (r-r_s)(r^2+l^2)}{\gamma^2 r -r+r_s}},\\
		\label{BetaDef2}
		\sin \beta &=&\frac{l }{r}\sqrt{\frac{r-r_s}{\gamma^2  r -r + r_s}},
	\end{eqnarray}	
\end{subequations}
where $l$ is presented in \eqref{Eq9.5}. The probability of transition from state $\ket{\alpha}$ at $r=\infty$ to state $\ket{\beta}$ at $r=\infty$ is given by \eqref{Eq5.27}:
\begin{equation}
	\label{Eq9.16}
	A_{\alpha\beta}= -\frac{i}{\hbar} \int_{0}^T d\tau \bra{\beta}V_{{eff}}\ket{\alpha} e^{-\frac{i\Delta E_{\alpha\beta}}{\hbar}\tau},
\end{equation}
where, 
\begin{equation}
	\label{Eq9.17}
	\Delta E_{\alpha\beta}= E_\alpha-E_\beta,
\end{equation}
and $T$ is the total (proper) time of flight of the hydrogen atom, which is given by,
\begin{equation}
	\label{Eq9.18}
	T =2 \int_{1}^{\infty}\frac{dr}{|\dot{r}|}.
\end{equation}
The (proper) time for the flight of the hydrogen atom from $\tau=0$ at $r=\infty$ is given by, 
\begin{equation}
	\label{Eq9.19}
	\tau(r) = \left\{
	\begin{array}{rcl}
		\frac{T}{2}- \int_1^r \frac{dr}{|\dot{r}|}&~~,~~& 0\leq \tau \leq\frac{T}{2}\\
		\\
		\frac{T}{2}+ \int_1^r \frac{dr}{|\dot{r}|}&,&\frac{T}{2}\leq \tau\leq T
	\end{array}
	\right.
\end{equation}
which can be substituted into \eqref{Eq9.16} to obtain,
\begin{equation}
	\label{Eq9.20}
	A_{\alpha\beta}= -\frac{2i}{\hbar}  e^{-\frac{i\Delta E_{\alpha\beta}}{2\hbar}T} \int_{1}^\infty \frac{d r}{|\dot{r}|} \bra{\beta}V_{{eff}}\ket{\alpha} \cos\left(\frac{\Delta E_{\alpha\beta}}{c\hbar} c\tau[r]\right),
\end{equation}
where $c$ is recovered, and $c\tau[r]$ is given by, 
\begin{equation}
	\label{Eq9.21}
	c\tau[r]=\int_1^r \frac{dr}{|\dot{r}|}.
\end{equation}
$\dot{r}$ is given by \eqref{Eq9.7}. We neglect the overall factor of $e^{-\frac{i\Delta E_{\alpha\beta}}{2\hbar}T}$ without loosing any generality. The effective potential presented in \eqref{Eq7.15} can be written as,
\begin{eqnarray}
	\label{Eq9.22}
	V_{{eff}} &=& V_{{eff}}^1 + V_{{eff}}^2 + V_{{eff}}^3,\\
	\label{Eq9.23}
	V_{{eff}}^1 &=& V_1~x^2(1+3 \cos 2\theta), \\
	\label{Eq9.24}
	V_{{eff}}^2 	&=& V_2~x^2\cos 2\phi \sin^2\theta,\\
	\label{Eq9.25}
	V_{{eff}}^3 	&=& V_3~ x^2\cos\phi \sin 2 \theta,
\end{eqnarray}
where, 
\begin{eqnarray}
	\label{Eq9.26}
	V_1&=& -\frac{m c^2 r_s }{32 r^3}(1+3 \cos 2\beta),\\
	\label{Eq9.27}
	V_2&=&-\frac{3m c^2 r_s}{16 r^3} \left(\frac{4 l^2}{r^2}+1 - \cos2\beta\right),\\
	\label{Eq9.28}
	V_3&=& +\frac{3m c^2 r_s}{4 r^3} \cosh\alpha \sin 2\beta.
\end{eqnarray}
Note that $V_1$ to $V_3$ depend on the position of hydrogen atom. They are not a function of the state of the hydrogen atom. We are looking for the transition from the ground state of the hydrogen atom. Therefore, we set,
\begin{equation}
	\label{Eq9.29}
	\ket{\alpha}= \ket{1,0,0}.
\end{equation}
We notice that,
\begin{eqnarray}
	\label{Eq9.30}
	\bra{n,\ell,m}V_{{eff}}^1\ket{1,0,0} &=& \delta^{m,0}\delta^{\ell,2} V_1 \bra{n,2,0}x^2\left(1+3 \cos 2\theta\right)\ket{1,0,0},\\
	\label{Eq9.31}
	\bra{n,\ell,m}V_{{eff}}^2\ket{1,0,0} &=& \delta^{m,\pm 2}\delta^{\ell,2} V_2 \bra{n,2,\pm 2}x^2\cos 2\phi \sin^2\theta\ket{1,0,0},\\
	\label{Eq9.32}
	\bra{n,\ell,m}V_{{eff}}^3\ket{1,0,0} &=& \delta^{m,\pm 1}\delta^{\ell,2} V_3 \bra{n,2,\pm 1}x^2\cos\phi \sin 2 \theta \ket{1,0,0},
\end{eqnarray}
where $\delta^{m,\ell}$ represents the Kronecker delta. We observe that the rules of transition from the state of $\ket{n,\ell,m}$ to state of $\ket{n',\ell',m'}$  for non-zero impact parameter of $\ell$ are,

	\begin{eqnarray}
		\label{Eq9.33}
		\ell-\ell'&=& \pm 2,\\
		m-m'&=& \pm 2, \pm 1, 0.
	\end{eqnarray}
The first excitation occurs for $n=3$, $\ell=2$, and $m=0,\pm1,\pm2$:
\begin{eqnarray}
	\label{Eq9.34}
	\bra{3,2,0}x^2(1+3 \cos 2\theta)\ket{1,0,0} &=&\frac{81}{32} \sqrt{\frac{3}{2}} a_0^2,\\	
	\label{Eq9.35}
	\bra{3,2,\pm 2}\left(x^2\cos 2\phi \sin^2\right)\theta\ket{1,0,0} &=& \frac{81}{512}a_0^2, \\
	\label{Eq9.36}
	\bra{3,2,\pm 1}\left(x^2\cos\phi \sin 2 \theta\right)\ket{1,0,0} &=& \frac{81}{128}a_0^2,
\end{eqnarray}
where $a_0$ is the reduced Bohr radius presented in \eqref{Eq7.6}.

\subsection{Probability of transition to {$\ket{3,2,0}$}}
The transition to $\ket{3,2,0}$ is due to $V_1$. Its amplitude  can be derived from \eqref{Eq9.20} using \eqref{Eq9.34}. It is given by, 
\begin{equation}
	\label{Eq9.37}
	A_{\ket{1,0,0}\to \ket{3,2,0}} =  -\frac{81\sqrt{3}m c^2 a_0^2}{512\sqrt{2} \hbar} r_s \int_1^{\infty} \frac{dr}{r^3|\dot{r}|}
	(1+3 \cos 2\beta)\cos\left(\frac{\Delta E}{c\hbar} c\tau[r]\right),
\end{equation}	
where $\Delta E= E_3-E_1$, $E_n$ are given in \eqref{Eq7.5}, and the overall phase of $-i e^{-\frac{i\Delta E}{2\hbar}T}$ is neglected. Using the values of $m, c, a_0$, and $\hbar$, one can get,
\begin{eqnarray}
	\label{Eq9.38}
	\frac{81\sqrt{3}m c^2 a_0^2}{512\sqrt{2} \hbar} &=& 26.55 a_0,\\
	\label{Eq9.39}
	\frac{\Delta E}{c\hbar}&=&  \frac{1}{308 a_0},
\end{eqnarray}
which can be employed in \eqref{Eq9.37},
\begin{eqnarray}
	\label{Eq9.40}
	A_{\ket{1,0,0}\to \ket{3,2,0}}  = -53.10\, a_0\, r_s \int_1^{\infty}dr \frac{\left(-1+3 \cos^2 \beta\right)}{|\dot{r}|r^3}
	\cos\left(\frac{c\tau[r]}{308 a_0} \right),
\end{eqnarray} 
where $\cos 2\beta$ is expressed in terms of $\cos \beta$.
$\cos \beta$, $\dot{r}$, $c\tau[r]$ and the unit of length are given in \eqref{Eq9.15},  \eqref{Eq9.7}, \eqref{Eq9.21}, and  by \eqref{Eq9.4}, respectively.
\begin{figure}[t]
	\begin{center}
		\includegraphics[width=0.8\textwidth]{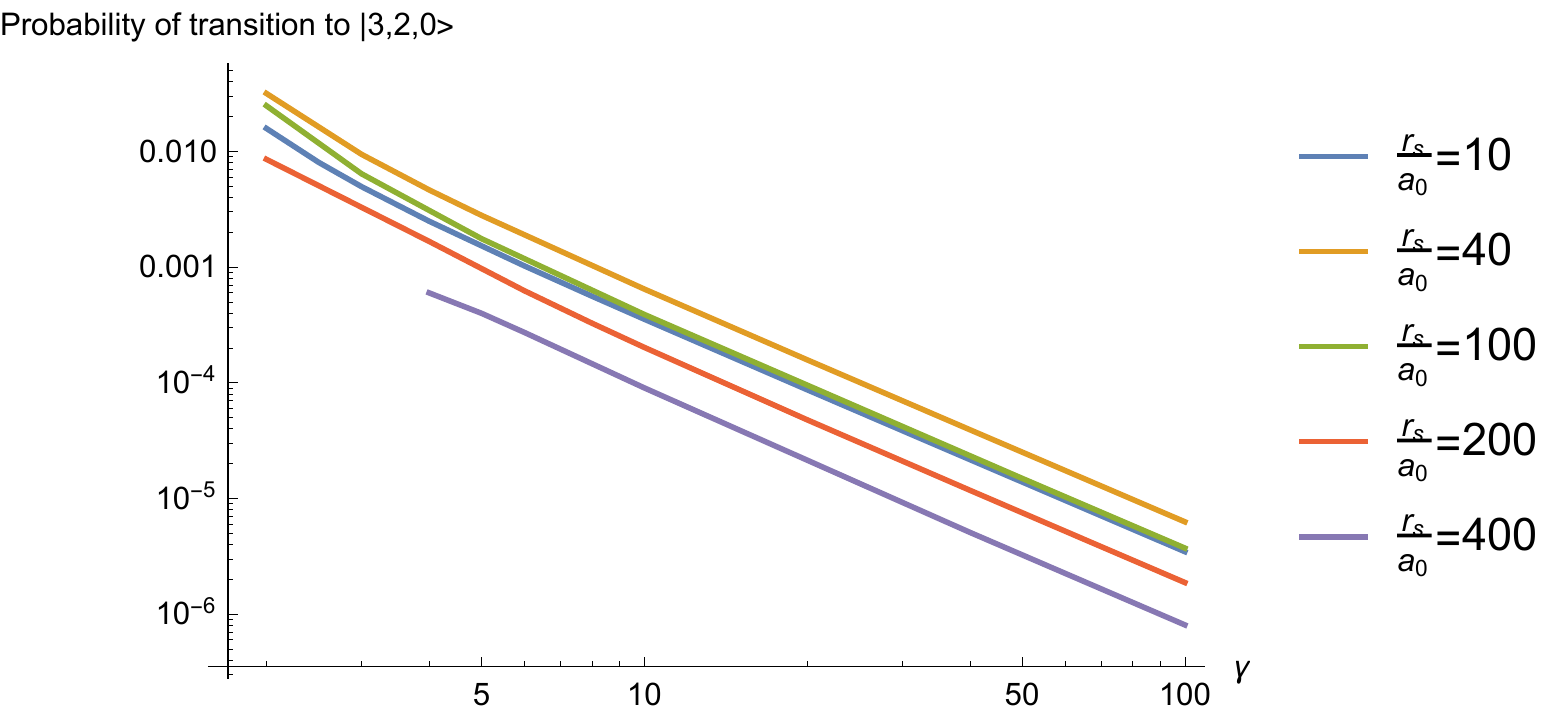}
	\end{center}
	\caption{Probability of transition to $\ket{3,2,0}$ in scattering by a black hole when the hydrogen atom reaches $10 a_0$ to the photon sphere, for a set of Schwarzschild radii, and in terms of the Lorentz factor of the hydrogen atom at the asymptotic infinity.}
	\label{fig:fig4}
\end{figure}

It is noticed that,
\begin{equation}
	\label{Eq9.41}
	\cos\left(\frac{c\tau[r]}{308 a_0} \right) \frac{1}{|\dot{r}|}= 308 a_0\frac{d}{dr}\sin\left(\frac{c\tau[r]}{308 a_0} \right),
\end{equation}
which can be utilized in \eqref{Eq9.38} to obtain,
\begin{eqnarray}
	\label{Eq9.42}
	A_{\ket{1,0,0}\to \ket{3,2,0}} \!=\!16335\, a_0^2  r_s \int_1^{\infty} dr 
	\sin\left(\frac{c\tau[r]}{308 a_0} \right)\frac{d}{dr}\left(\frac{-1+3 \cos^2 \beta}{r^3}\right),
\end{eqnarray} 
where integration by parts is performed. The transition amplitude of \eqref{Eq9.42} is written in the length unit given by \eqref{Eq9.4}. $r_{{min}}$ can be recovered by, 
\begin{eqnarray}
	\label{Eq9.43}
	r_s &\to & \frac{r_s}{r_{{min}}},\\
	\label{Eq9.44}
	a_0 & \to & \frac{a_0}{r_{{min}}}.
\end{eqnarray} 
For any value of $r_{{min}}$, $r_s$, and $\gamma$, Eq. \eqref{Eq9.42} can numerically be computed. For example, when the hydrogen atom reaches the minimum distance of $d a_0$ to the photon sphere, it holds that,
\begin{equation}
	\label{Eq9.45}
	r_{{min}}= \frac{3}{2} r_s + d a_0.
\end{equation}
Defining $r_s= \zeta a_0$ and utilizing \eqref{Eq9.4} then gives,
\begin{eqnarray}
	\label{Eq9.46}
	r_s &=&\frac{2 \zeta }{2 d + 3 \zeta},\\
	\label{Eq9.47}
	a_0 &=&\frac{2}{2 d + 3 \zeta}, 
\end{eqnarray}
which can be substituted into \eqref{Eq9.42} to numerically calculate the amplitude for any value of $d$ and $\zeta$. Figure \eqref{fig:fig4} depicts the probability of transition to $\ket{3,2,0}$ for the hydrogen atom at $d=10$, and a set of Schwarzschild radii in terms of the Lorentz factor. We observe a simple behaviour at large $\gamma$. We would like to look at the large $\gamma$ limit of \eqref{Eq9.42} for arbitrary value of $r_s$ and $r_{{min}}$. 

The large $\gamma$ limit says that,
\begin{equation}
	\label{Eq9.48}
	\frac{d}{dr}\left(\frac{-1+3 \cos^2 \beta}{r^3}\right) = \frac{3 \left(-2 r^3 (r_s-1)-5 r+6 r_s\right)}{r^7 (r_s-1)}+O\left(\frac{1}{\gamma^2 }\right).
\end{equation}
Using the large $\gamma$ limit of $\dot{r}$ defined in \eqref{Eq9.7}, \eqref{Eq9.21} yields, 
\begin{equation}
	\label{Eq9.49}
	c\tau[r]= \int_1^\gamma \frac{dr}{|\dot{r}|}= \frac{1}{\gamma} g(r) + O\left(\frac{1}{\gamma^2}\right),
\end{equation}
where $g(r)$ does not depends on $\gamma$ and is given by,
\begin{equation}
	\label{Eq9.50}
	g(r)= \int_1^r \frac{dr}{\sqrt{1-\frac{r-r_s}{r^3(1-r_s)}}}.
\end{equation}
\begin{figure}[t]
	\begin{center}
		\includegraphics{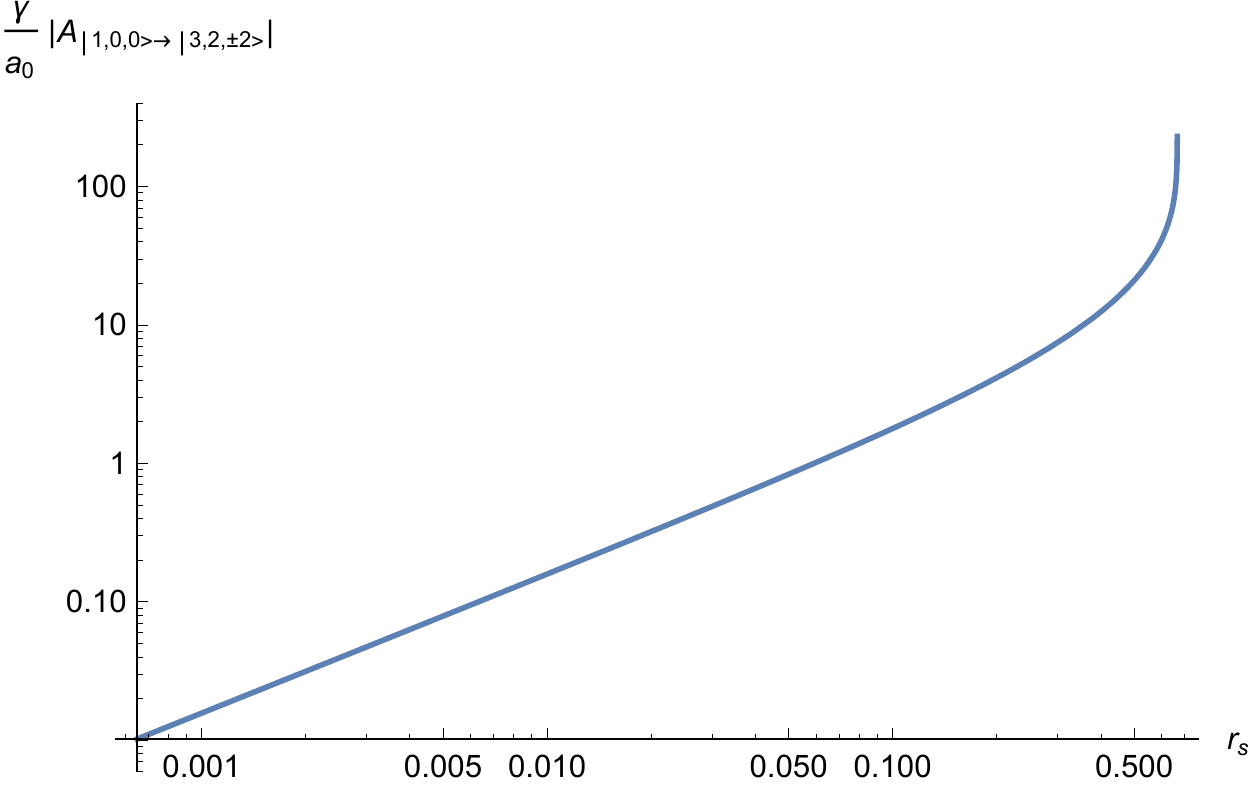}
	\end{center}
	\caption{The amplitude of transition for large $\gamma$ in terms of the Schwarzschild radius. There exists an essential singularity at the photon's sphere, at $r_s=\frac{2}{3}$.}
	\label{fig:fig5}
\end{figure}
The large $\gamma$ limit of \eqref{Eq9.42} is thus given by, 
\begin{eqnarray}
	\label{Eq9.51}
	\frac{A_{\ket{1,0,0}\to \ket{3,2,0}}}{16335\,a_0^2}= r_s \int_1^{\infty} dr \sin \left(\frac{g(r)}{308 a_0 \gamma}\right) \frac{3 \left(-2 r^3 (r_s-1)-5 r+6 r_s\right)}{r^7 (r_s-1)},\nonumber\\
\end{eqnarray} 
where $O\left(\frac{1}{\gamma^2}\right)$ is ignored. 
The fall off of the integrand for $r\to \infty$ is $r^6$. So focusing at $r\approx 1$ gives a good approximation to the integral. Near $r=1$, the sine factor in the integrand can be approximated by,
\begin{equation}
	\label{Eq9.52}
	\sin \left(\frac{g(r)}{308 a_0 \gamma}\right) \approx \frac{g(r)}{308 a_0 \gamma}.
\end{equation} 
The above expression can be employed to simplify the amplitude to,
\begin{eqnarray}
	\label{Eq9.53}
	A_{\ket{1,0,0}\to \ket{3,2,0}}  &=&53.10\, \frac{a_0  r_s}{\gamma} \int_1^{\infty} dr 
	g(r)\frac{3 \left(-2 r^3 (r_s-1)-5 r+6 r_s\right)}{r^7 (r_s-1)}\nonumber\\
	&+& O\left(\frac{1}{\gamma^2}\right)\!,
\end{eqnarray}  
where $g(r)$ is given in \eqref{Eq9.50}. Integration by parts simplifies \eqref{Eq9.53} to, 
\begin{eqnarray}
	\label{Eq9.54}
	\frac{A_{\ket{1,0,0}\to \ket{3,2,0}}}{53.10}  &=& \frac{a_0  r_s}{\gamma} \int_1^{\infty} dr 
	g'(r) \int_0^r dx \frac{3 \left(-2 x^3 (r_s-1)-5 x+6 r_s\right)}{x^7 (r_s-1)}\nonumber\\
	&+& O\left(\frac{1}{\gamma^2}\right)\!.
\end{eqnarray}  
The integral in the integrand can be performed, and \eqref{Eq9.50} can be used to write,
\begin{eqnarray}
	\label{Eq9.55}
	A_{\ket{1,0,0}\to \ket{3,2,0}}  &=&-53.10 \frac{a_0  r_s}{\gamma} \int_1^{\infty} dr 
	\frac{3 r + 2 r^3 (-1 + r_s) - 3 r_s}{r^{\frac{9}{2}}\sqrt{(1-r_s)^2 r^3-(r-r_s)(1-r_s)}}  \nonumber\\
	&+& O\left(\frac{1}{\gamma^2}\right)\!.
\end{eqnarray}  
\begin{figure}[t]
	\begin{center}
		\includegraphics[width=0.8\textwidth]{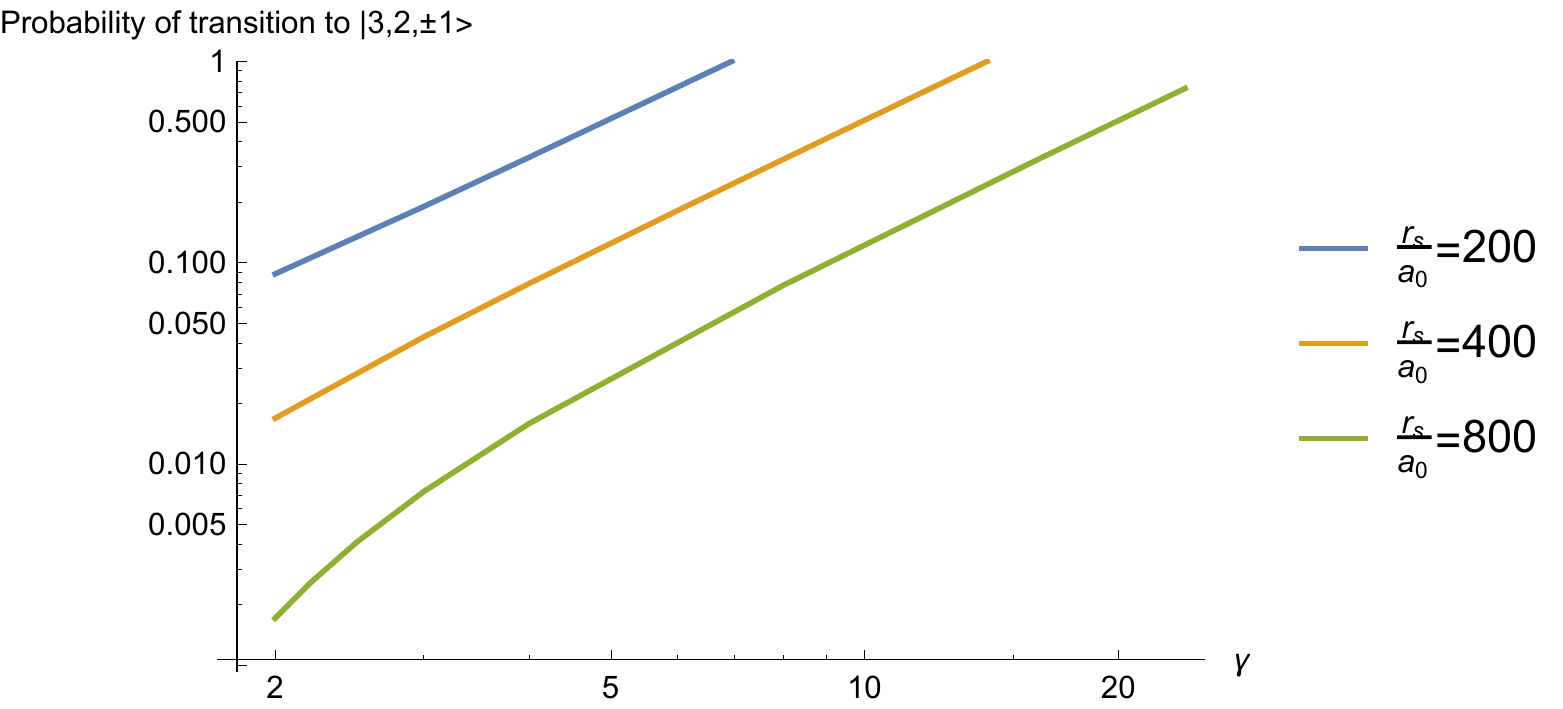}
	\end{center}
	\caption{The probability of transition to $\ket{3,2,\pm 1}$ for $r_{{min}}=2 r_s$, and for a set of $r_s$ in terms of $\gamma$.}
	\label{fig:fig6}
\end{figure}
Equation \eqref{Eq9.55} has an essential singularity at the photon's sphere for $r_3=\frac{2}{3}$. The singularity can be best seen by plotting the amplitude in terms of $r_s$, as depicted in Fig. \ref{fig:fig5}. Let it be emphasized that the unit of length is defined by \eqref{Eq9.4}, where $r_{{min}}$ is the minimum distance of the hydrogen atom to the black hole during its `journey'. 
We notice that the integral in \eqref{Eq9.55} is finite and non-zero. Therefore, the amplitude of transition to $\ket{3,2,0}$ vanishes in the large limit of $\gamma$. This means that, for a very large value of $\gamma$, the hydrogen atom either falls inside the black hole, or when it is deflected by the black hole, it is not excited to $\ket{3,2,0}$. 

\subsection{Probability of transition to {$\ket{3,2,\pm 1}$}
}
Transition to $\ket{3,2,\pm 1}$ is due to $V_2$.  Its amplitude  can be derived from \eqref{Eq9.20} by utilizing \eqref{Eq9.31} and  \eqref{Eq9.35}. It is given by, 
\begin{eqnarray}
	\label{Eq9.56}
	A_{\ket{1,0,0}\to \ket{3,2,\pm 1}}\!= \!-\frac{243m c^2 a_0^2}{4096 \hbar} r_s \int_1^{\infty}\!\! \frac{dr}{r^3|\dot{r}|}
	\left(\frac{4 l^2}{r^2}+1 - \cos2\beta\right)\cos\left(\frac{\Delta E}{c\hbar} c\tau[r]\right)\!, \nonumber\\
\end{eqnarray}	
where $\Delta E= E_3-E_1$, $E_n$ are given in \eqref{Eq7.5}, and the overall phase of $-i e^{-\frac{i\Delta E}{2\hbar}T}$ is neglected.
\begin{figure}[t]
	\begin{center}
		\includegraphics[width=0.8\textwidth]{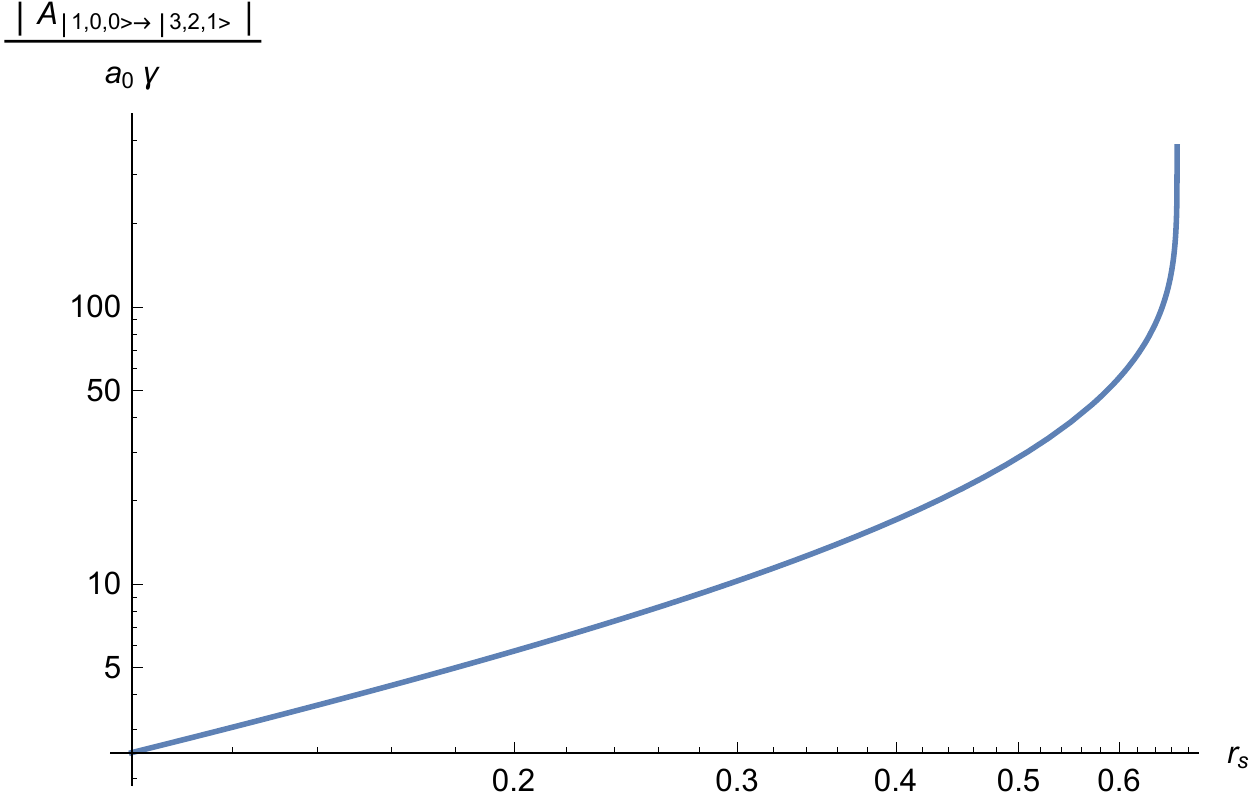}
	\end{center}
	\caption{Amplitude of transition to $\ket{3,2,\pm1}$ for large $\gamma$ in terms of the Schwarzschild radius of the black hole. There exists an essential singularity at the photon's sphere, $r_s=\frac{2}{3}$.}
	\label{fig:fig7}
\end{figure}
Utilizing the values of $m, c, a_0$ and $\hbar$ 
,
\begin{eqnarray}
	\label{Eq9.57}
	\frac{243m c^2 a_0^2}{4096 \hbar} &=& 8.13 a_0,\\
	\label{Eq9.58}
	\frac{\Delta E}{c\hbar}&=&  \frac{1}{308 a_0},
\end{eqnarray}
which can be employed in \eqref{Eq9.56}, 
\begin{eqnarray}
	\label{Eq9.59}
	A_{\ket{1,0,0}\to \ket{3,2,\pm 1}}  = -8.13 a_0 r_s \int_1^{\infty}dr \,\frac{\frac{4 l^2}{r^2}+2\sin^2\beta}{|\dot{r}|r^3}
	\cos\left(\frac{c\tau[r]}{308 a_0} \right),
\end{eqnarray} 
where we used the identity $1- \cos 2\beta=2 \sin^2 \beta$. Using Eq. \eqref{BetaDef2}, one obtains,
\begin{eqnarray}
	\label{Eq9.60}
	A_{\ket{1,0,0}\to \ket{3,2,\pm 1}} \!= \!-16.26\, a_0 r_s l^2 \int_1^{\infty}
	\frac{dr}{|\dot{r}|r^5}
	\left(2 + \frac{r-r_s}{r_s+r(\gamma^2-1)}\right) 
	\cos\left(\frac{c\tau[r]}{308 a_0} \right). \nonumber\\
\end{eqnarray} 
Using \eqref{Eq9.41} and performing integration by parts, the above expression can be further simplified to,
\begin{eqnarray}
	\label{Eq9.61}
	\frac{A_{\ket{1,0,0}\to \ket{3,2,\pm 1}}}{5008.08\, a_0^2 r_s l^2}  =  \int_1^{\infty}\!
	dr	\sin\left(\frac{c\tau[r]}{308 a_0} \right)
	\frac{d}{dr} \left(\frac{2}{r^5} + \frac{r-r_s}{(r_s+r(\gamma^2-1))r^5}\right),\nonumber\\
\end{eqnarray} 
which can numerically be calculated for given values of $r_s, a_0$, and $\gamma$. Figure \ref{fig:fig6} depicts the probability for $r_{{min}}=2 r_s$, and for various values of $r_s$. It is observed that the transition probability has a simple behaviour at large $\gamma$.
\begin{figure}[t]
	\begin{center}
		\includegraphics[width=0.8\textwidth]{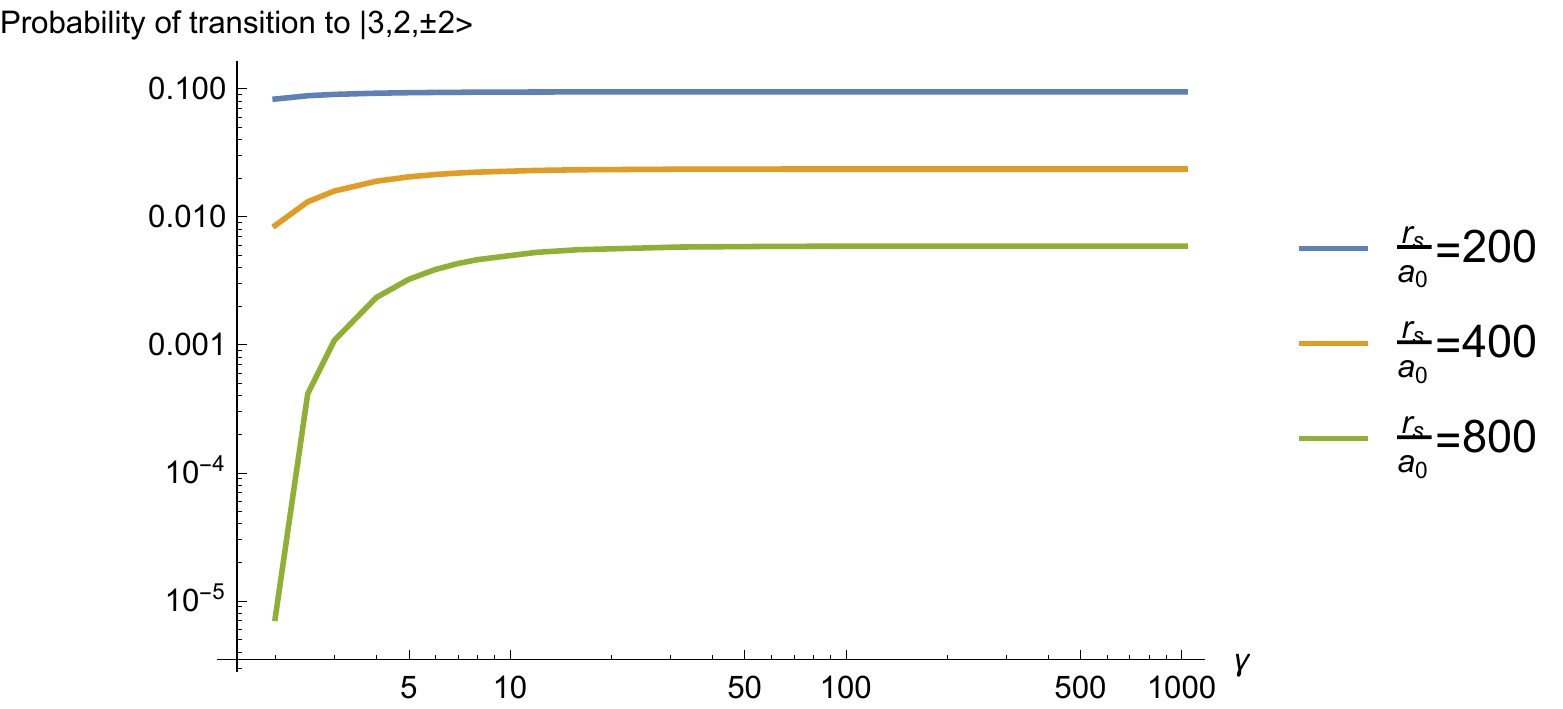}
	\end{center}
	\caption{The probability of transition to $\ket{3,2,\pm 2}$ for $r_{{min}}=2 r_s$, and for a set of $r_s$ in terms of $\gamma$.}
	\label{fig:fig8}
\end{figure}

In the large $\gamma$ limit, we can use \eqref{Eq9.49} and \eqref{Eq9.52} to simplify the amplitude to, 
\begin{eqnarray}
	\label{Eq9.62}
	A_{\ket{1,0,0}\to \ket{3,2,\pm 1}}  = 16.26 \frac{a_0 r_s l^2}{ \gamma} \int_1^{\infty}
	dr g(r) 
	\frac{d}{dr}\left(\frac{2}{r^5} - O\left(\frac{1}{\gamma^2}\right)\right),
\end{eqnarray} 
where $l^2$ is presented in \eqref{Eq9.5} and, in the large $\gamma$ limit, is simplified to,  
\begin{equation}
	\label{Eq9.63}
	l^2= \frac{\gamma^2}{1-r_s}.
\end{equation}
So, 
\begin{eqnarray}
	\label{Eq9.64}
	A_{\ket{1,0,0}\to \ket{3,2,\pm 1}}  = -162.6 \frac{a_0 r_s \gamma}{ 1-r_s} \int_1^{\infty}
	dr \frac{g(r)}{r^6},
\end{eqnarray} 
where $g(r)$ is given in \eqref{Eq9.50}. Utilizing $\frac{1}{r^6}= -\frac{1}{5}\partial_r\frac{1}{r^5}$ and performing an integration by part yields, 
\begin{eqnarray}
	\label{Eq9.65}
	A_{\ket{1,0,0}\to \ket{3,2,\pm 1}}  = 32.52 \frac{a_0 r_s \gamma}{ 1-r_s} \int_1^{\infty}
	\frac{dr}{r^5}\frac{1}{\sqrt{1-\frac{r-r_s}{r^3(1-r_s)}}}.
\end{eqnarray} 
Figure \ref{fig:fig7} shows the absolute amplitude in terms of $r_s$. There exists an essential singularity at the photon's sphere, $r_s=\frac{2}{3}$. The essential singularity points that, when the hydrogen atom approaches the photon's sphere, it will be easily excited to $\ket{3,2,\pm 1}$.  The transition to $\ket{3,2,1}$ occurs for sufficiently large $\gamma$ at any $r_s$.

\subsection{Probability of transition to {$\ket{3,2,\pm 2}$}}
The transition to $\ket{3,2,\pm 2}$ is due to $V_3$.  Its amplitude can be derived from \eqref{Eq9.20} by employing \eqref{Eq9.32} and  \eqref{Eq9.36}. It is given by,
\begin{equation}
	\label{Eq9.66}
	A_{\ket{1,0,0}\to \ket{3,2,\pm 2}} =  -\frac{243m c^2 a_0^2}{256 \hbar} r_s \int_1^{\infty} \frac{dr}{r^3|\dot{r}|}
	\cosh \alpha \sin 2\beta \cos(\frac{\Delta E}{c\hbar} c\tau[r]),
\end{equation}	
where $\Delta E= E_3-E_1$, $E_n$ are given in \eqref{Eq7.5}, and the overall phase of $-i e^{-\frac{i\Delta E}{2\hbar}T}$ is neglected. Utilizing the values of $m, c, a_0$ and $\hbar$ 
,
\begin{eqnarray}
	\label{Eq9.67}
	\frac{243m c^2 a_0^2}{256 \hbar} &=&130.08 a_0,\\
	\label{Eq9.68}
	\frac{\Delta E}{c\hbar}&=&  \frac{1}{308 a_0},
\end{eqnarray}
which can be employed in \eqref{Eq9.56}, 
\begin{eqnarray}
	\label{Eq9.69}
	A_{\ket{1,0,0}\to \ket{3,2,\pm 2}}  =-130.08  a_0 r_s \int_1^{\infty} \frac{dr}{r^3|\dot{r}|}
	\cosh \alpha \sin 2\beta \cos\left(\frac{c\tau[r]}{308 a_0}\right),
\end{eqnarray} 
\begin{figure}[t]
	\begin{center}
		\includegraphics[width=0.8\textwidth]{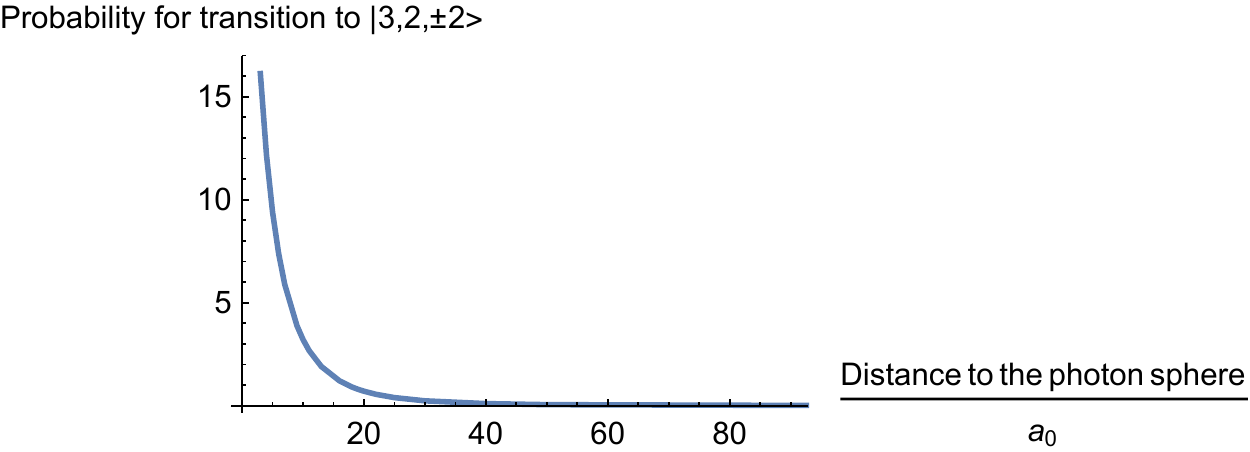}
	\end{center}
	\caption{The probability of transition to $\ket{3,2,\pm 2}$ for $\gamma=1000$ and $r_s=10 a_0$, versus the minimum distance to the photon sphere.}
	\label{fig:fig9}
\end{figure}
Using \eqref{Eq9.41} and performing integration by part gives,
\begin{eqnarray}
	\label{Eq9.70}
	A_{\ket{1,0,0}\to \ket{3,2,\pm 2}}  =40064.6  a_0^2 r_s \int_1^{\infty} dr \,
	\sin\left(\frac{c\tau[r]}{308 a_0}\right) \frac{d}{dr}\left(\frac{\cosh \alpha \sin 2\beta }{r^3}\right).\nonumber\\
\end{eqnarray}
Figure \ref{fig:fig8} depicts the transition probability for $r_{{min}}=2 r_s$ and for a set of $\frac{r_s}{a_0}$. It is observed that the large $\gamma$ limit of the amplitude possesses a simple behaviour.  
The large $\gamma$ limit yields, 
\begin{eqnarray}
	\label{Eq9.71}
	\frac{d}{dr}\left(\frac{\cosh \alpha \sin 2\beta }{r^3}\right) =\frac{\gamma  \left(-8 r^3 (r_s-1)-10 r+11 r_s\right)}{r^{13/2} (r_s-1) \sqrt{r^3 (-(r_s-1))-r+r_s}}+O\left(\frac{1}{\gamma^0 }\right).\nonumber\\
\end{eqnarray}
So \eqref{Eq9.52} can be used to simplify the amplitude in large $\gamma$ to\footnote{Notice that  \eqref{Eq9.13} demands $r_s\leq \frac{2}{3}$. The integrand, thus, is always real.}, 
\begin{eqnarray}
	\label{Eq9.72}
	A_{\ket{1,0,0}\to \ket{3,2,\pm 2}}  &=&-\frac{130.08  a_0 r_s}{1-r_s} \int_1^{\infty} \frac{dr}{r^{13/2}} 
	\frac{ g(r) \left(8 r^3 (1-r_s)-10 r+11 r_s\right)}{ \sqrt{r^3 (1-r_s)-r+r_s}}\nonumber\\
	&+& O\left(\frac{1}{\gamma}\right),
\end{eqnarray}
where $g(r)$ is given in \eqref{Eq9.50}, and the integral is a finite non-zero number for $r_s <\frac{2}{3}$. It, however, has an essential singularity at the photon's sphere at $r_s=\frac{2}{3}$. The singularity can be seen in Fig. \ref{fig:fig9} where the probability of transition is depicted for $\gamma=1000$, $r_s=10 a_0$ versus the minimum distance to the photon's sphere.

\section{Discussions and Conclusion}
\label{Section10}
The effect of the curved space-time geometry on static hydrogen atoms  \cite{Yu:2007wv,Zhou:2012eb,Cheng:2019tnk,Zhu:2008bd} and the change in the spectrum of a static hydrogen atom in a curved space-time geometry has been studied  \cite{Parker:1982nk,Parker:1980kw} in the literature. Here, we have computed the effect of the curved space-time geometry on a non-static hydrogen atom. We have considered the low-energy physics in the effective field theory approach to QFT in a general curved space-time geometry admitting an asymptotic (flat) infinity. We have considered a localised quantum system that moves along a time-like geodesic. We have reviewed and utilised  Fermi coordinates along the geodesic to describe the space-time geometry around the geodesic wherein  the metric is the Minkowski metric corrected by the Riemann tensor evaluated on the geodesic. We have calculated the leading correction by the curved space-time geometry to the Schr\"odinger equation. We have shown that the components of the Riemann tensor introduce a time dependent perturbative potential which distorts the wave-function, causing excitation as the quantum system moves along the time-like geodesic. 

Through direct computation, we have illustrated how the curvature of space-time geometry causes an excitation for the hydrogen atom when it falls in or is scattered by a Schwarzschild black hole. We have shown that the excitation is enhanced when the hydrogen atom possesses a speed with respect to the background space-time geometry. The enhancement suggests that storing local information in a ultra-relativistic probe may not be trivial as a tiny change in the background space-time geometry  may induce a quantum excitation in the stored memory. 

We have shown that a freely falling hydrogen atom  gets excited by the background black hole by direct computation. The excited states then emit a photon by the spontaneous emission and decay into the ground state.  Emitted photons generally escape to the asymptotic infinity, and can be detected.  We tend to argue that the energy of these photons comes from the curved space-time geometry, so they extract energy from the black hole, causing the black hole  to gradually decay into the flat space-time geometry.  Though this decay turns out to be negligible and not detectable by the current technology, it is interesting that the interaction of a quantum system with the black hole induces a new decay channel for the black hole, a channel in addition to Hawking radiation \cite{Hawking:1974rv}. This perhaps adds to the information paradox.

Last but not least, we would like to point that it would be interesting to compute how the emission from hydrogen atom predicted here would the constrain distribution of the primordial black holes. For such a purpose, one should assume a profile for the distribution of the primordial black holes and their velocity, compute the electromagnetic radiation emitted due to the interaction of clouds of the hydrogen atoms with the primordial black holes, and find how CMB observation would constrain the distribution of the primordial black hole. 

\section*{Acknowledgments}
This work was supported by the High Throughput and Secure Networks Challenge Program at the National Research Council of Canada, the Canada Research Chairs (CRC) and Canada First Research Excellence Fund (CFREF) Program, and Joint Centre for Extreme Photonics (JCEP). We thank Alicia Sit for reading the manuscript and providing comments and feedback. We thank Mattias Blau and Niayesh Afshordi for comments, feedback and discussions.

\providecommand{\newblock}{}

\end{document}